\documentclass[longbibliography,prl,aps,twocolumn,showpacs,floatfix]{revtex4-2}
\usepackage{amsfonts}
\usepackage{amssymb}
\usepackage{hyperref}
\usepackage{graphicx}
\usepackage{dcolumn}
\usepackage{bm,amsmath,verbatim}
\usepackage{mathrsfs}
\usepackage{color}
\usepackage{lineno}
\usepackage{dsfont}
\usepackage{amsmath,amssymb,amsthm}

\usepackage[T1]{fontenc}
\usepackage[latin9]{inputenc}
\usepackage{booktabs}

\hypersetup{colorlinks,
linkcolor=blue,          citecolor=blue,        filecolor=blue,      urlcolor=blue           }
\def\jr{{j}}
\def\jrp{{j'}}
\def\lr{{l}}
\def\thr{{\theta}}
\def\kr{{k}}

\def\be{\begin{equation}}
\def\ee{\end{equation}}
\def\bea{\begin{eqnarray}}
\def\eea{\end{eqnarray}}
\def\bse{\begin{subequations}}
\def\ese{\end{subequations}}

\def\be{\begin{eqnarray}}
\def\ee{\end{eqnarray}}

\begin{document}

\setlength\columnsep{25pt}
\title{Anomalous quantized nonlinear soliton pumping}
\author{Yu-Liang Tao$^{1}$}
\author{Jiong-Hao Wang$^{1}$}
\author{Yong Xu$^{1,2}$}
\email{yongxuphy@tsinghua.edu.cn}
\email{yongxuphy@mail.tsinghua.edu.cn}
\affiliation{$^{1}$Center for Quantum Information, IIIS, Tsinghua University, Beijing 100084, People's Republic of China}
\affiliation{$^{2}$Hefei National Laboratory, Hefei 230088, People's Republic of China}

\begin{abstract}
	It has recently been theoretically predicted and experimentally observed that a soliton resulting from nonlinearity can
	be pumped across an integer or fractional number of unit cells as a system parameter is slowly varied over a pump period.
	Nonlinear soliton pumping is now understood as the flow of instantaneous Wannier functions,
	ruling out the possibility of pumping a soliton across a nonzero number of unit cells
	over one cycle when a corresponding Wannier function does not exhibit any flow, i.e., when the corresponding 
	Bloch band that the soliton bifurcates from is topologically trivial. 
	Here we surprisingly find an anomalous nonlinear soliton pump where the displacement of a soliton over one cycle 
	differs from the Chern number of the Bloch band from which the soliton comes. 
	We show that this anomalous behavior arises from a transition of a soliton
	between different Wannier functions by passing through an intersite-soliton (or dipole-soliton) state.
	Furthermore, we find a nonlinearity-induced integer 
	quantized pump of a soliton, allowing a soliton to travel across one unit cell during a pump period, even when the 
	corresponding band is topologically trivial.
	Our results open the door to studying nonlinearity-induced pumping of solitons.    
\end{abstract}

\maketitle

\section{Introduction}
Thouless pumping, 
which plays an important role in understanding the quantum Hall effect,
is a phenomenon where quantized transport arises through the 
slow periodic variation of a Hamiltonian parameter~\cite{thoulessPRB1983,citro2023thouless}. 
The transport requires the complete filling of an energy band so that it is 
dictated by the Chern number of the occupied band with the varying parameter playing the role of 
an additional momentum. So far, Thouless pumps have been experimentally observed in photonic systems~\cite{krausPRL2012,zilberbergNat2018,cerjanLSA2020},
cold atom systems~\cite{nakajimaNP2016,lohseNP2016,lohseNat2018,nakajimaNP2021,walterNP2023} 
and other systems~\cite{maPRL2018,chengPRL2020,fedorovaNC2020}.  

Nonlinearity is a natural feature in various systems, including photonic systems~\cite{christodoulidesOL1988,eisenbergPRL1998,fleischerNat2003,christodoulidesNat2003,kivsharbook2003,ledererPR2008discrete,kevrekidisBook2009} and 
Bose-Einstein condensates (BECs)~\cite{brazhnyiMPLB2004,morschRMP2006,dauxoisbook2006,BEC_book_2008,chinRMP2010,busch2000motion}. It leads to many intriguing phenomena, such as 
solitons, which are wave packets that travel without changing their shape. 
Nonlinearity also gives rise to topology-related phenomena, 
such as topological bulk~\cite{lumerPRL2013,mukherjeeSci2020} and edge solitons~\cite{ablowitzPRA2014,leykamPRL2016,mukherjeePRX2021,tao2020hinge}, 
as well as nonlinearity-induced topological insulators~\cite{maczewskySci2020,soneNP2024}.
In particular, it has recently been found that a soliton arising from nonlinearity can 
exhibit quantized transport when an underlying Hamiltonian parameter is slowly 
tuned periodically (called nonlinear Thouless pumping)~\cite{jurgensenNat2021,jurgensenPRL2022,fuPRL2022,mostaanNC2022,
	fu2022twoD,tuloupNJP2023,szameit2024discrete}. 
Later, a quantized fractional Thouless pumping of solitons is found by considering 
a soliton bifurcating from multiple bands~\cite{jurgensenNP2023}. 
The integer and fractional quantized soliton pumping are now understood 
as the flow of the instantaneous one-band and maximally localized multi-band 
Wannier functions~\cite{jurgensenPRL2022,fuPRL2022,mostaanNC2022,jurgensenNP2023}, respectively.
For the former, the displacement of a soliton is determined by 
the Chern number of the Bloch band of the linear Hamiltonian of the system~\cite{jurgensenPRL2022,fuPRL2022,mostaanNC2022}.
If this mechanism always worked, we would not observe the occurrence that a soliton's displacement
is different from this Chern number.
In other words, a soliton would not undergo transport when a linear Hamiltonian
is topologically trivial.

Here we surprisingly find an anomalous nonlinear soliton pump that can exhibit the pumping of a 
soliton across zero, two, or three unit cells
over one cycle, while the Chern number of the band, from which the soliton comes, is $C=-1$. 
Our result is thus beyond the previous nonlinear soliton pump where the displacement is identical
to the Chern number. 
We show that this anomalous behavior arises from a transition of a soliton 
between different Wannier functions by passing through an intersite-soliton (or dipole-soliton) state.
Leveraging the discovery of the new nonlinear pump, we construct an alternative nonlinear model, 
demonstrating the occurrence of quantized nonlinear pumping even when the Chern number of the band of 
the linear Hamiltonian vanishes (in other words, the linear Thouless pumping 
does not happen in this system over one cycle). 
To associate a Chern number to the anomalous nonlinear pumping, 
we study a spatial supercell containing multiple unit cells to ensure the well localization of a soliton within the supercell. 
Subsequently, we calculate the Chern number of a modulated Hamiltonian 
involving effects of the soliton solution, 
finding it to be identical to the displacements.
Finally, we show the emergence of anomalous nonlinear pumping in a nonlinear continuous model, 
which can be experimentally realized in cold atom systems.   

\section{Results}
\subsection{Novel mechanism}
\begin{figure}
	\includegraphics[width=1\linewidth]{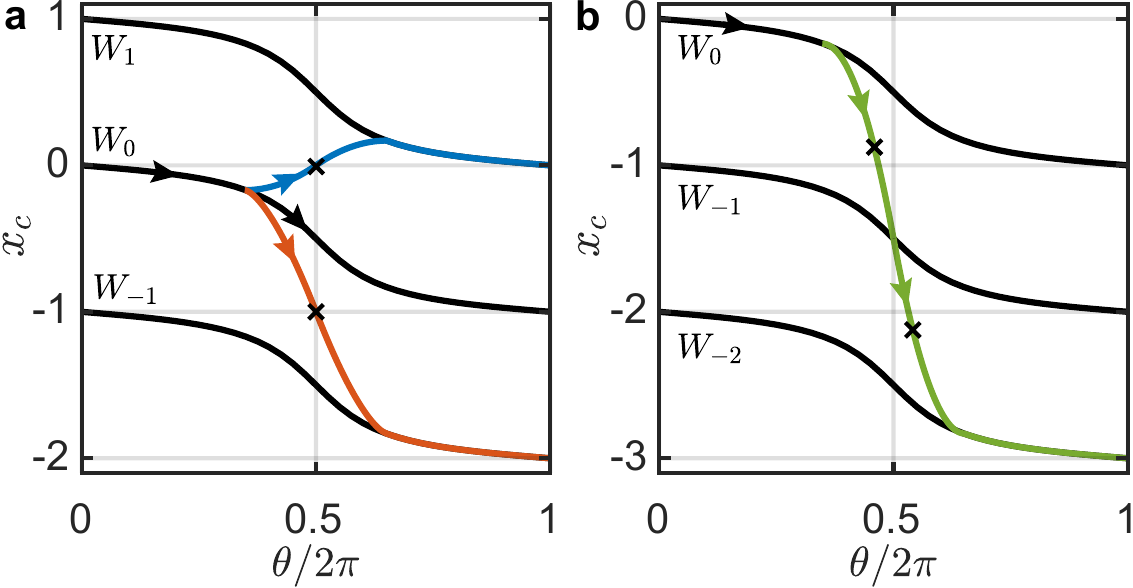}
	\caption{\textbf{Schematic illustration of how anomalous nonlinear pumping arises.} 
		The black lines represent the center-of-mass position of Wannier functions with respect to $\theta$. 
		When a soliton follows a black line, normal nonlinear pumping appears.
		In contrast, if a soliton transitions between Wannier functions by
		passing through intersite solitons, whose center-of-mass positions are marked by 
		diagonal crosses, anomalous nonlinear pumping occurs. For example, 
		following the blue and red lines in \textbf{a} results in displacements of $0$
		and $-2$, respectively, while following the green line in \textbf{b} results in
		a displacement of $-3$.
	}
	\label{fign1}
\end{figure}

We start by showing in general how the anomalous 
nonlinear soliton pumping occurs. Consider 
a one-dimensional (1D) non-interacting Hamiltonian $H^{\mathrm{lin}}(\theta)$ with translation symmetry. 
The Hamiltonian is controlled by a system parameter $\theta$ which
varies from $0$ to $2\pi$, satisfying $H^{\mathrm{lin}}(0)=H^{\mathrm{lin}}(2\pi)$.
The Hamiltonian at each $\theta$ has Bloch states characterized by momentum $k$.
We focus on an isolated Bloch band associated with a set of Wannier
functions denoted by $W_l(j,\theta)$, where $l$ is an integer denoting the lattice vector in 1D,
and $j$ is the spatial coordinate.
Without 
loss of generality, we assume that the Chern number of this band with 
respect to momentum $k$ and parameter $\theta$ is $-1$ (although our theory is applicable to 
other Chern numbers as well).
Consequently, the center-of-mass position (referred to as the Wannier center) of $W_l(j,\theta)$
changes by $-1$ as $\theta$ varies from $0$ to $2\pi$, assuming that 
the length of a unit cell is $1$, as shown by the black lines in Fig.~\ref{fign1}.
Prior research indicates that, in the presence of nonlinearity, soliton solutions that resemble 
Wannier functions (referred to as onsite solitons) exist, such that the soliton undergoes a 
displacement equal to the Chern number
over one pump period~\cite{jurgensenNat2021,jurgensenPRL2022,fuPRL2022,mostaanNC2022}.
Therefore, in this case, the soliton experiences a displacement of $-1$ over one cycle.

In addition to the onsite-soliton solution, 
previous studies have also investigated intersite (or dipole) soliton solutions~\cite{kivsharbook2003,ledererPR2008discrete,kevrekidisBook2009}.
We define an ideal intersite soliton as an equal superposition of two neighboring Wannier functions, 
e.g., $[W_0(j,\theta)+W_1(j,\theta)]/\sqrt{2}$.
Note that while the intersite soliton might be reminiscent of the doublon~\cite{doublons}, they are 
completely different. 
Although intersite solitons themselves have been extensively studied~\cite{dipole,Morandotti1999PRL,Kapitula2001,
	Yuri2003PRL,Hadzievski2004PRL,Lewenstein2004PRA,Pelinovsky2005,Lluis2005PRL,Melvin2006PRL}, 
in the context of nonlinear soliton pumping, 
prior studies have found that an intersite soliton is always unstable~\cite{jurgensenPRL2022}, 
thus precluding the stable evolution between onsite and intersite solitons,
as a system parameter is varied.
In addition, in a single-band, single-component setting, since 
a Wannier function is always predominantly localized at a single site, 
no system parameter can induce its displacement. 
Therefore, to achieve anomalous nonlinearity pumping, one 
must consider a system with multiple degrees of freedom per unit cell. 

We now show that a soliton can transition from one Wannier function to another
by passing through an intersite-soliton state, leading to the anomalous nonlinear pumping.
Specifically, consider a system where, at $\theta=0$, a soliton solution is $W_0(j,0)$, 
with a center-of-mass position of $0$.
If the soliton remains in the Wannier state as $\theta$ increases, 
its center-of-mass will follow the black trajectory, as illustrated in Fig.~\ref{fign1}, consistent
with previous reports. 
However, we find that near $\theta=\pi$, stable intersite solitons exist,
e.g., $\left[W_0(j,\pi)+W_1(j,\pi)\right]/\sqrt{2}$ and 
$\left[W_{0}(j,\pi)+W_{-1}(j,\pi)\right]/\sqrt{2}$, 
with their center-of-mass positions marked by diagonal crosses in
Fig.~\ref{fign1}a. 
Through such an intersite soliton, the soliton can transition from 
$W_0(j,0)$ into either $W_1(j,2\pi)$ or $W_{-1}(j,2\pi)$,
following the blue or red trajectory and yielding quantized transports of 
$0$ and $-2$, respectively, 
both of which are distinct from the Chern number. 
Although the former case is similar 
to the trapped soliton for strong nonlinearity~\cite{jurgensenNat2021,fuPRL2022}, the mechanism is
completely different (see Supplementary Note 1 for details).
The trapped phenomenon arises from all band contributions due to strong 
nonlinearity, leading to zero Chern number~\cite{jurgensenNP2023}.
In contrast, in our case, the soliton always occupies one single band.

Moreover, as shown in Fig.~\ref{fign1}b, a soliton can first evolve from 
$W_0(j,0)$ to $W_{-1}(j,\pi)$ through an intersite-soliton state
formed by $W_0$ and $W_{-1}$ and subsequently evolve to 
$W_{-2}(j,2\pi)$ through an intersite-soliton state formed by $W_{-1}$ and $W_{-2}$.
This process results in a displacement of $-3$.
These anomalous transport outcomes 
do not align with the Chern number of the linear band.
The theory suggests the potential for various anomalous results; 
however, it is crucial that the soliton solutions remain stable throughout the entire period.

\begin{figure*}
	\includegraphics[width=1\linewidth]{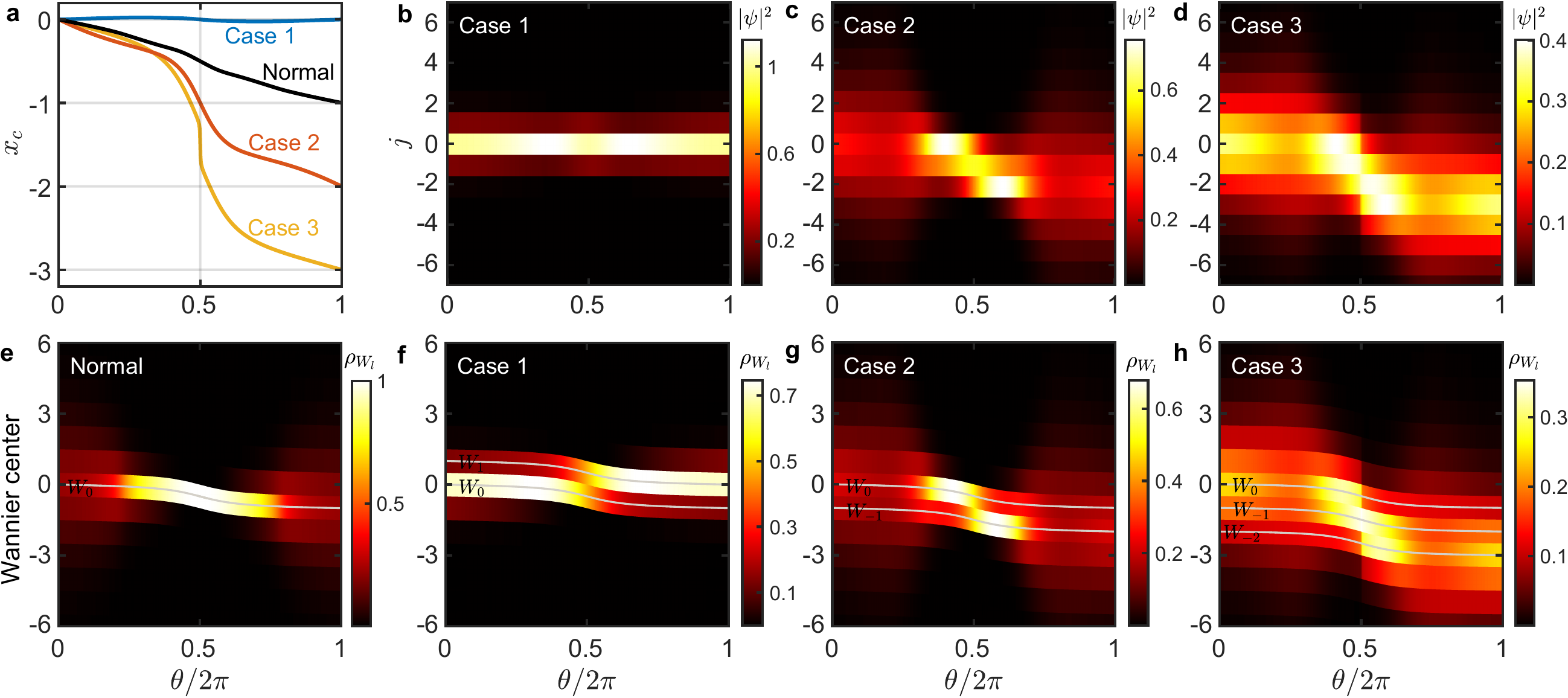}
	\caption{\textbf{Anomalous nonlinear soliton pumping in the discrete nonlinear model.}
		\textbf{a,} One-cycle trajectory of center-of-mass positions of instantaneous solitons
		coming from the lowest band of the linear Hamiltonian in Eq.~(\ref{DNSE})
		as a system parameter $\theta$ varies.
		It plots three anomalous cases corresponding to displacements of $0$ (case 1), 
		$-2$ (case 2), and $-3$ (case 3), described by the blue ($g=-1$, $g_{12}=0$, and $m_0=1$), 
		red ($g=1$, $g_{12}=0$, and $m_0=1$), and gold lines ($g=1$, $g_{12}=0$, and $m_0=1.3$), 
		respectively, in contrast to the normal case with a displacement of $-1$
		described by the black line ($g=g_{12}=m_0=1$).
		The associated density distributions $|\psi_j|^2=\sum_{\sigma}|\psi_{\sigma j}|^2$ of the solitons
		for the three anomalous cases are plotted in \textbf{b--d}. 
		\textbf{e--h,} 
		Occupations $\rho_{W_l}(\theta)$ of instantaneous solitons on 
		Wannier functions $W_l(\theta)$ of the lowest band along the Wannier centers as a function of $\theta$
		for the four cases. See also Supplementary Note 3 for more information.
		Here, we set $N=1.45$.
	}
	\label{fign2}
\end{figure*}

\subsection{Discrete nonlinear model}
To demonstrate that the anomalous nonlinear soliton 
pumping can arise in concrete models, we study the following dimensionless 
discrete nonlinear Schr\"odinger equation:
{\begin{eqnarray}\label{DNSE}
		i\frac{\partial}{\partial t}\psi_{\sigma \jr}= \sum_{\sigma^\prime \jrp} 
		H_{\sigma \jr, \sigma^\prime \jrp}^{\textrm{lin}}(\theta) \psi_{\sigma^\prime \jrp}
		+V_{\sigma j}(\psi_{\sigma j},\psi_{\overline{\sigma} j}) \psi_{\sigma j},
\end{eqnarray}}
where $\psi_{\sigma \jr}$ is the wavefunction of the $\sigma$th component ($\sigma=1,2$) 
in the $\jr$th unit cell at time $t$ ($t$ is the propagation distance in photonic systems),
$H^{\textrm{lin}}$ is the linear tight-binding Hamiltonian that depends on a system parameter $\theta$, 
$V_{\sigma j}(\psi_{\sigma j},\psi_{\overline{\sigma} j})=g|\psi_{\sigma j}|^2+g_{12} |\psi_{\overline{\sigma} j}|^2$
with $g$ and $g_{12}$ being 
nonlinear coefficients, and $\overline{\sigma}=(\sigma \text{ mod } 2)+1$
labels the other component. The norm of the wavefunction $N=\sum_{\sigma \jr} |\psi_{\sigma \jr}|^2$ is preserved during 
the evolution in this equation. 
We consider a linear Hamiltonian which reads in momentum space
\begin{equation} \label{HLin}
	H^{\textrm{lin}}(k)=(m_z+J_1 \cos k) \sigma_z+J_1^\prime \sin k \sigma_y+ J_2  \sigma_x,
\end{equation} 
where $\sigma_\nu$ ($\nu=x,y,z$) are Pauli matrices, and $m_z=m_0+\cos \theta$ with $m_0$ being a real system parameter. 
We set $J_1=J_1^\prime =1$ and $J_2=\sin \theta$. The parameter $\theta$
is slowly varied as time evolves, realizing a periodic change of the linear Hamiltonian.
This model can be transformed to the Rice-Mele model through a unitary transformation
realized by $U=e^{-i\pi\sigma_y/4}$,
and it clearly realizes the Chern band~\cite{qiPRB2008} if we view $\theta$ as the other momentum $k_y$ besides $k$.
The Chern number of the lowest band with respect to $k$ and $\theta$ is $C=1$ 
when $-2<m_0<0$, $C=-1$ when $0<m_0<2$, and $C=0$ otherwise.
For linear Thouless pumping, particles are required to fill an entire band so that  
the cloud of particles will travel across the lattice distance identical to the Chern number as we 
slowly tune $\theta$ from $0$ to $2\pi$~\cite{thoulessPRB1983}. 
Note that this tight-binding model can be experimentally 
implemented in ultracold atomic gases as proposed in Ref.~\cite{yangPRB2018} 
with $g$ and $g_{12}$ describing the intraspecies and interspecies 
interactions, respectively 
(see the section on nonlinear continuous model).

To demonstrate the anomalous transport behavior 
in this nonlinear model, we calculate stable instantaneous soliton solutions 
$\chi_{\sigma j}(\theta)$
that come from the lowest band with the 
Chern number $C=-1$ for the instantaneous nonlinear Hamiltonian at each $\theta$ 
using the Newton's method; the solution is related to $\psi_{\sigma j}$ through
$\psi_{\sigma j}=e^{-i\mu(\theta) t} \chi_{\sigma j}(\theta)$, where 
$\mu(\theta)$ is the chemical potential.
This approach is justified because, in the adiabatic limit---where $\theta$
is varied very slowly---the time evolution of a soliton follows the instantaneous eigenstate of the 
instantaneous nonlinear Hamiltonian at time $t$~\cite{kivsharRMP1989,bandPRA2002,bandPRA2002_2,liuPRL2003,wuPRL2005}.
We have directly computed the time evolution based on Eq.~(\ref{DNSE}) when $\theta$
is varied sufficiently slowly and found that the results closely 
resemble the instantaneous solutions. 
Additionally, we have performed a stability analysis to confirm that the solved nonlinear instantaneous 
soliton solutions remain stable at each $\theta$
(see also Supplementary Note 2).

\begin{figure*}
	\includegraphics[width=1\linewidth]{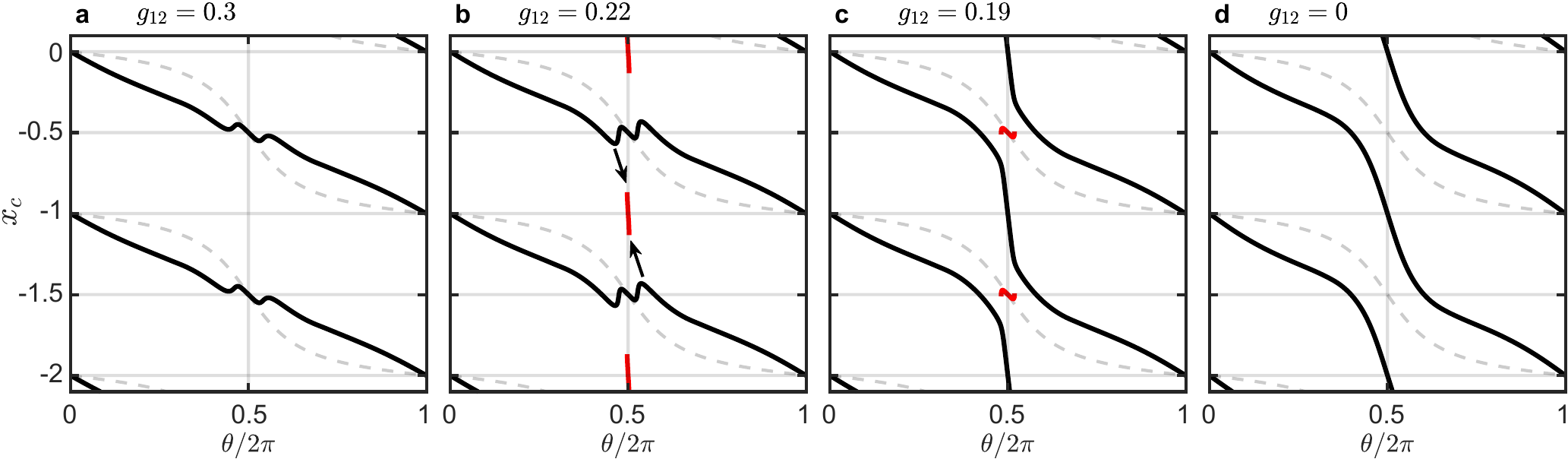}
	\caption{\textbf{Transition between normal and anomalous nonlinear soliton pumping.} \textbf{a--d,} 
		Center-of-mass positions of stable instantaneous solitons (black and red lines) and 
		Wannier functions (dashed grey lines) with respect to a system 
		parameter $\theta$ over one cycle for different $g_{12}$. 
		The anomalous nonlinear pumping arises 
		due to appearance of intersite soliton solutions
		around $\theta=\pi$ (see red lines in \textbf{b}). 
		See Supplementary Note 5 for stability analysis at $\theta=\pi$.
		Here, we set $m_0=g=1$ and $N=1.45$.}
	\label{fign3}
\end{figure*}

Figure~\ref{fign2}a illustrates that when $g_{12}=1$, 
the soliton's center of mass travels across the 
lattice by $-1$ over one period, which is equal to the Chern number 
of the lowest band. Here the center of mass is defined as
$x_c(\theta)=\sum_{\sigma j}j |\psi_{\sigma j}|^2/N$, and the soliton displacement 
over one cycle is given by $x_c(2\pi)-x_c(0)$. 
This result is consistent with the previous findings~\cite{jurgensenNat2021,jurgensenPRL2022,fuPRL2022,mostaanNC2022}.
Surprisingly, we observe that when $g_{12}=0$, 
the soliton's displacement changes to $0$, $-2$, or $-3$, 
which differ from the Chern number of the lowest band
(see also Figs.~\ref{fign2}b--d for evolution of the soliton profiles). 
Here, we set $g=1$
(see Supplementary Note 4 for discussion on
effects of $g$ on soliton's width). 
The anomalous nonlinear pumping can occur over a wide parameter region.
For example, given nonlinear parameters $|g|=1$ and $N=1.45$, 
case 1 occurs in the range $-0.66 \le g_{12} \le 0$ for $m_0=1$, case 2 in 
$0 \le g_{12} \le 0.19$ for $m_0=1$, and case 3 in $0 \le g_{12} \le 0.01$ for $m_0=1.3$. 
Here, we focus on negative $g_{12}$ for case 1 and positive $g_{12}$ for case 2 and case 3. 
Although the parameter regime for case 3 with $m_0=1.3$ is relatively narrow, 
the phase diagram with respect to $m_0$ and $g_{12}$ exhibits a wide parameter 
regime (see Supplementary Fig. 5 in Supplementary Note 4).

To confirm that the anomalous nonlinear pumping
results from a transition to an intersite soliton,
we expand the instantaneous soliton solution $\chi_{\sigma j}(\theta)$ in terms
of Wannier functions with the occupation on the $l$th Wannier function being given by
$\rho_{W_l}(\theta)=|\sum_{\sigma j}W_{\sigma l}^*(j,\theta)\chi_{\sigma j}(\theta)|^2/N$,
where $W_{\sigma l}(j,\theta)$ is the $l$th Wannier function for the $\sigma$th component 
from the lowest Bloch band. For notation simplicity, we denote 
$W_{\sigma l}$ by $W_l$, without specifying its component $\sigma$.
The bottom row in Fig.~\ref{fign2} illustrates $\rho_{W_l}(\theta)$ along the Wannier centers. 
In the normal case, the soliton is dominated by a Wannier function $W_0$ 
as $\theta$ varies (see Fig.~\ref{fign2}e), thereby 
following the trajectory of the Wannier function.
However, for the anomalous case, such as 
in Fig.~\ref{fign2}g, a soliton initially dominated by $W_0$ evolves 
into an ideal intersite soliton composed of $W_0$ and $W_{-1}$ 
at $\theta=\pi$ (see derivation of intersite soliton solutions in Supplementary Note 5). 
Subsequently, the soliton becomes dominated by $W_{-1}$. 
This result is consistent with the theory presented in the previous section.
Similar transitions occur in the other two cases.

We proceed to examine how 
nonlinear pumping transitions to the anomalous one.
To illustrate this, we plot in Fig.~\ref{fign3} the center-of-mass position of stable nonlinear 
eigenstates of the instantaneous 
nonlinear Hamiltonian as a function of $\theta$ for different values of $g_{12}$.
Notably, 
when $g_{12}=0.3$,  
the center-of-mass positions of a soliton and the Wannier centers exhibit clear differences 
for most values of $\theta$, but they coincide at $\theta=0$ and $\theta=\pi$.
In fact, at $\theta=\pi$, the soliton solution is precisely the Wannier function 
multiplied by $\sqrt{N}$ when $m_0=1$ (see Supplementary Note 5). 
However, as $g_{12}$ decreases, 
a stable intersite soliton solution emerges at $\theta=\pi$, localized around $j=l$ with $l$ being an integer, 
differing from the Wannier center, which is located at $l+1/2$. With a further decrease in $g_{12}$,
we observe that the new branch of solutions becomes longer,
eventually connecting to the original soliton solutions (represented by black lines). Consequently,  
when $g_{12}<0.2$, a soliton becomes displaced by $-2$ during adiabatic 
evolution.

\subsection{Nonlinearity-induced soliton pumping}
\begin{figure}
	\includegraphics[width=1\linewidth]{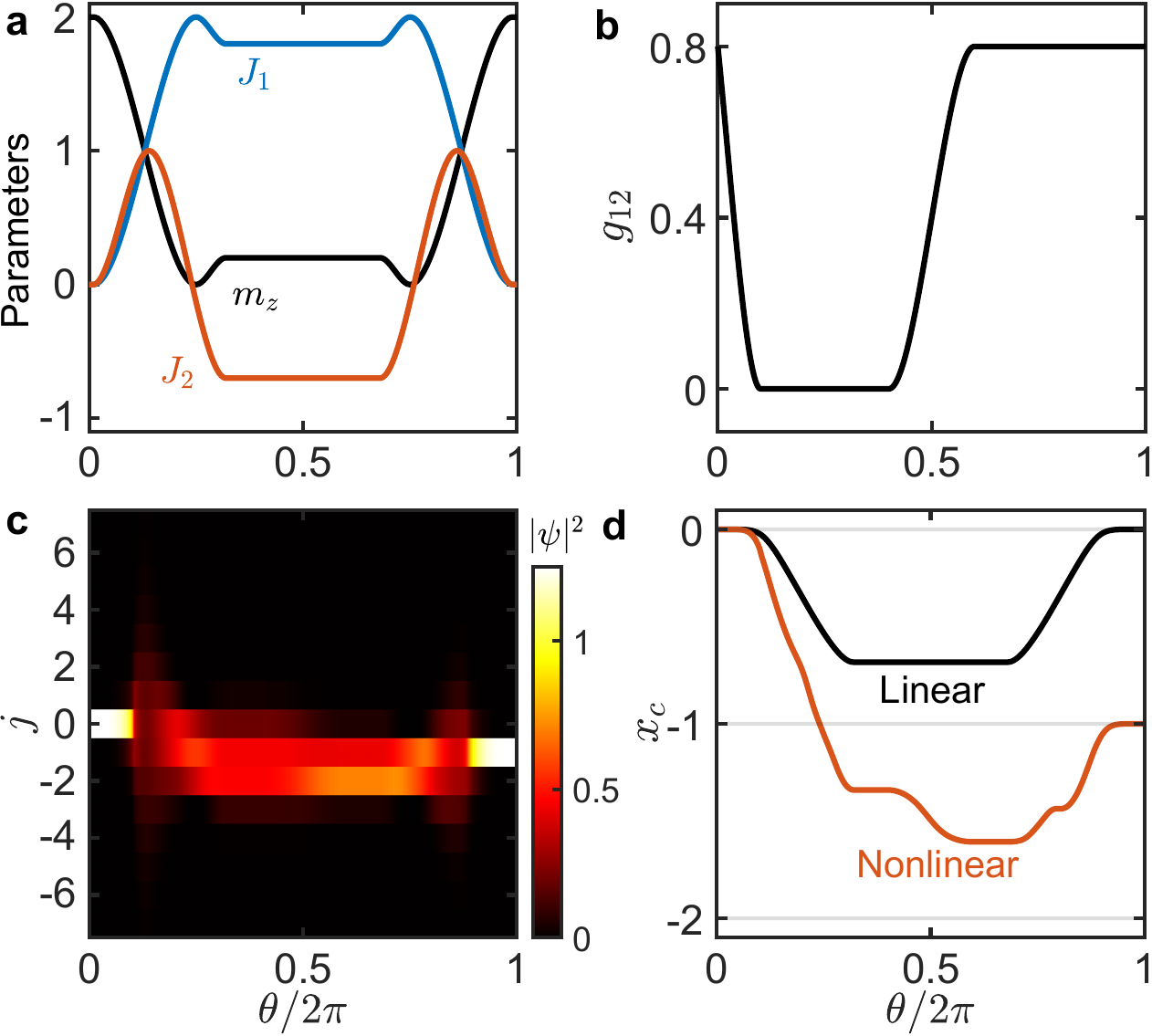}
	\caption{\textbf{Nonlinearity-induced soliton pumping.} \textbf{a,b,} 
		Illustration of how the parameters $m_z(\theta)$, $J_1(\theta)$, $J_2(\theta)$ and 
		$g_{12}(\theta)$ are varied with respect to $\theta$ 
		in order to induce a nonlinear pumping for a topologically trivial band. 
		\textbf{c,} One-cycle evolution 
		of the density distribution $|\psi_\jr|^2$ of instantaneous solitons 
		bifurcating from the lowest band. \textbf{d,} 
		The evolution of center-of-mass positions of instantaneous solitons (red line) and the corresponding 
		Wannier functions (black line) with respect to $\theta$ over one cycle. Here, we set $N=1.25$ and $g=1$.}
	\label{fign4}
\end{figure}

We have demonstrated the breakdown of the correspondence between displacement and 
the Chern number for a nonlinear pump, which naturally leads to a question whether a
soliton can be pumped when the Chern number of the linear band is zero. 
Our answer is affirmative. To illustrate this, we again consider the linear Hamiltonian in Eq.~(\ref{HLin}). 
Now, we vary the system parameters
$m_z$, $J_1=J_1^\prime$ and $J_2$ with respect to $\theta$.
For simplicity, we impose the conditions that all these parameters 
are symmetric about $\theta=\pi$ as illustrated in Fig.~\ref{fign4}a.
Consequently, the Hamiltonian $H^{\textrm{lin}}(k,\theta)$ and
the Zak phase $\gamma_n(\theta)$ also have this symmetry;
$\gamma_n(\theta)=i\int_0^{2\pi}{\rm d}k\langle u_{n,k}|\partial_ku_{n,k}\rangle$ 
is the Zak phase of the 1D system 
at a fixed $\theta$ with $|u_{n,k}\rangle$ being the $n$th eigenstate of $H^{\textrm{lin}}(k,\theta)$.
The Chern number for each band is thus zero so that the linear Thouless pumping cannot occur.

To achieve nonlinearity-induced pumping of solitons, we simultaneously adjust $g_{12}$, 
as shown in Fig.~\ref{fign4}b
(the interaction can be controlled via Feshbach resonances in ultracold atomic gases~\cite{chinRMP2010}).
Specifically, we initially reduce $g_{12}$ to zero gradually, which facilitates the soliton's 
travel over a longer distance. Subsequently, we slowly increase $g_{12}$ back to its original
value to prevent the soliton from returning to its initial position. In Fig.~\ref{fign4}c, 
we plot the density distribution of nonlinear eigenstates of the instantaneous nonlinear Hamiltonian 
as $\theta$ varies from $0$ to $2\pi$, remarkably demonstrating that the soliton is displaced by $-1$ unit cell.
This effect is further evidenced in Fig.~\ref{fign4}d, which depicts the evolution of center-of-mass 
positions of instantaneous soliton solutions as $\theta$ changes.
This observation starkly contrasts with the behavior of the Wannier center (or the Zak phase), which 
returns to its starting value after one complete cycle. Hence, we successfully achieve pumping of a 
soliton through the influence of nonlinearity.

\begin{figure*}
	\includegraphics[width = 1\linewidth]{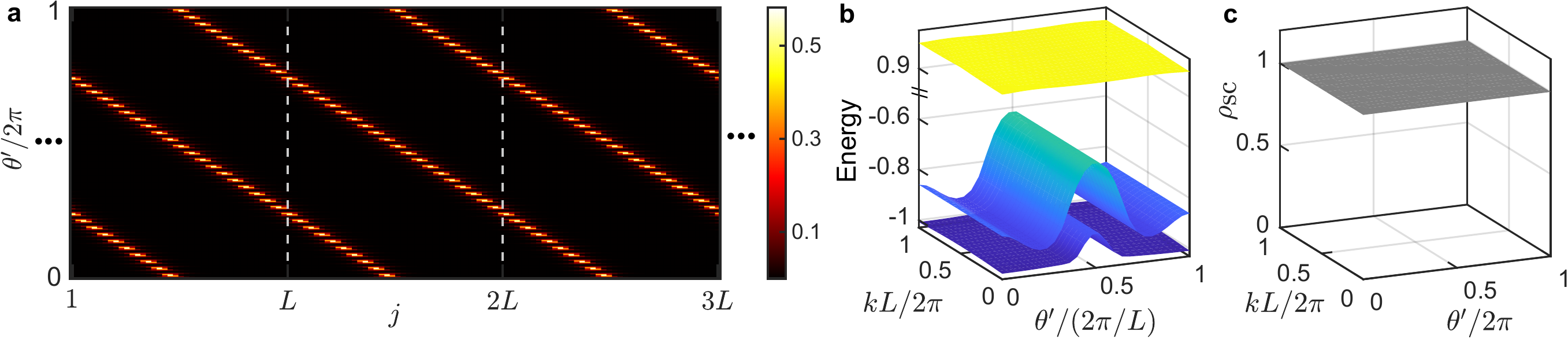}
	\caption{\textbf{Supercell method.}
		\textbf{a,} The effective potential generated by a soliton array with one soliton per supercell 
		(three supercells are explicitly shown, each containing $L=40$ sites) as 
		$\theta^\prime=\theta/L$ varies from $0$ to $2\pi$. 
		The potential is periodic with period $L$. Only the potential felt by the first component is 
		plotted, and the potential for the other component is similar. 
		\textbf{b,} 
		Energy spectra of the modulated linear Hamiltonian $H^{\textrm{sc}}$ containing the linear potentials in \textbf{a} 
		with respect to momentum $k$ and 
		$\theta^\prime$. Given that the spectra are periodic functions of $\theta^\prime$ with a period of $2\pi/L$, 
		we only plot part of the spectrum with $\theta^\prime\in[0,2\pi/L]$. Here, 
		we show three Bloch bands around $\mu(\theta)$ and the second band corresponds 
		to the soliton wavefunction $\psi_{\sigma j}^{\textrm{s}}(\theta)$.
		\textbf{c,} The normalized projection of a soliton wavefunction $\psi_{\sigma j}^{\textrm{s}}$
		on the second band in \textbf{b}, 
		$
		\rho_{\textrm{sc} }(k,\theta^\prime)=
		\frac{L_{\textrm{sc} }}{N} |\sum_{\sigma j}\varphi_{k ,\sigma, j}(\theta^\prime)^*\psi_{\sigma j}^{\textrm{s}}(\theta^\prime)|^2,
		$
		where $L_{\textrm{sc}}$ is the number of supercells, 
		$\varphi_{k,\sigma,j}(\theta^\prime)$ is a Bloch state at momentum $k \in [0,2\pi/L]$ in the corresponding band of the Hamiltonian 
		$H^{\textrm{sc}}$.
		Here, we consider case 2 in Fig.~\ref{fign2} with $N=1.45$, $m_0=1$, $g_{12}=0$, and $g=1$.
	}
	\label{fign5}
\end{figure*}

To associate a Chern number to the nonlinear pump itself,
we consider a modulated linear Hamiltonian in a supercell consisting of $L$ unit cells
incorporating the effects of solitons~\cite{jurgensenNat2021}. 
To establish the validity of this method, consider a soliton solution 
$\psi_{\sigma j}^{\textrm{s}}(\theta)$ that satisfies Eq.~(\ref{DNSE}). This solution is 
used to define a new linear Hamiltonian at each $\theta$ as
\begin{equation}
	H^{\prime}_{\sigma j,\sigma^\prime j^\prime}(\theta) := H^{\textrm{lin}}_{\sigma j,\sigma^\prime j^\prime}(\theta)
	+V_{\sigma j}(\psi_{\sigma j}^{\textrm{s}}(\theta),\psi_{\overline{\sigma} j}^{\textrm{s}}(\theta))
	\delta_{j j^\prime}\delta_{\sigma \sigma^\prime},
\end{equation}
where $V_{\sigma j}(\psi_{\sigma j}^{\textrm{s}}(\theta),\psi_{\overline{\sigma} j}^{\textrm{s}}(\theta))$ is the effective potential generated by the soliton.
We then study the time-dependent linear equation,
$i\partial_t \Psi_{\sigma j}=\sum_{\sigma^\prime j^\prime} 
H^{\prime}_{\sigma j,\sigma^\prime j^\prime}(\theta) \Psi_{\sigma^\prime j^\prime}
$,
where $H^\prime$ is independent of $\Psi$. Clearly, $\psi_{\sigma j}^{\textrm{s}}(\theta)$ 
is a solution to this linear equation, thus converting a nonlinear problem to a linear one. 
However, $H^\prime$ does not preserve translational symmetry due to the soliton-induced defect potential. 
To restore the symmetry, we partition the lattice into sections $\dots,[1,L],[L,2L],\dots$, 
each called a supercell, composed of $L$ unit cells.
$L$ is sufficiently large so that a soliton is well localized within a supercell at $\theta=0$. 
At this $\theta$, a soliton exists within each supercell, as illustrated in Fig.~\ref{fign5}a,
making the effective potential caused by the solitons periodic with a period of $L$. 
Since the solitons are far away from each other, the soliton array
should also be the solution to Eq.~(\ref{DNSE}). 
Similar to $H^\prime(\theta)$, we define a linear Hamiltonian $H^{\textrm{sc}}(\theta)$
using the effective potential of the soliton array at each $\theta$,
which respects the translational symmetry, i.e., 
$H^{\textrm{sc}}_{\sigma j+L,\sigma^\prime j^\prime+L}(\theta)=
H^{\textrm{sc}}_{\sigma j,\sigma^\prime j^\prime}(\theta)$.

Our numerical calculations indicate that the soliton in each supercell 
closely resembles a Wannier function of the Hamiltonian $H^{\textrm{sc}}(\theta)$.
This is evidenced by the uniform projections of the soliton wavefunction onto 
the second band in Fig.~\ref{fign5}b, as shown in Fig.~\ref{fign5}c, 
indicating that the soliton wavefunctions are equal superpositions of 
all the Bloch states in this band, and are thus Wannier functions.
Since the displacement of a Wannier function over a pump period is determined by 
the Chern number $C_{\textrm{sc}}$ of this band of 
$H^{\textrm{sc}}$, the soliton's displacement is thus dictated by this Chern number. 
To see this, we introduce a new parameter $\theta^\prime=\theta/L$ such that 
$H^{\textrm{sc}}(\theta^\prime+2\pi)=H^{\textrm{sc}}(\theta^\prime)$. 
As $\theta^\prime$ varies from $0$ to $2\pi$, the Wannier function travels 
by $C_{\textrm{sc}}$ supercells, equivalent to $C_{\textrm{sc}} L$ original unit cells, 
indicating that the average displacement over a cycle of $\theta$ is $C_{\textrm{sc}}$. 
Our calculations reveal that for normal nonlinear pumping, $C_{\textrm{sc}}$
is equal to the Chern number $C$ of the corresponding Bloch band of $H^{\textrm{lin}}$;
however, they differ in the anomalous one.
For example, in the anomalous cases shown 
in Fig.~\ref{fign2}, $C_{\textrm{sc}}$ can be $0$, $-2$, or $-3$ , while $C=-1$.
In Fig.~\ref{fign4}, $C_{\textrm{sc}}=-1$ while $C=0$.
In addition, we find that $C_{\textrm{sc}}$ changes from $-1$ to $-2$
as $g_{12}$ decreases across $0.2$ in Fig.~\ref{fign3}.

\subsection{Continuous nonlinear model}
To implement the tight-binding model in Eq.~(\ref{HLin})
in ultracold atomic gases, we consider the following continuous model: 
\begin{equation} \label{Hc}
	H_{\textrm{c} }^{\textrm{lin} }(x,\theta)=H_0(x)+h_z(\theta)\sigma_z+V_{\textrm{so} }(x,\theta)\sigma_x,
\end{equation}
where the spin is encoded in two hyperfine states of an atom. 
Note that similar models have been proposed for realizing linear Thouless pumping in both
two-dimensional and three-dimensional cold atom systems~\cite{yangPRB2018}.
In this Hamiltonian,  
$H_0(x)=\frac{p_x^2}{2m}-V_x \cos^2 (k_R x)$, where
$p_x=-i\hbar \partial_x$ is the momentum operator, $m$ is the mass of atoms, 
and $V_x$ is the 
strength of optical lattices generated by lasers with wavevector $k_R$.
In addition, $h_z$ denotes the strength of the Zeeman field corresponding to the detuning of the Raman lasers, and
$V_{ \textrm{so} }(x,\theta)=V_s \sin(k_R x)+V_c\cos(k_R x)\sin(\theta)$. This potential can be implemented using two pairs of 
Raman lasers, as described in Ref.~\cite{yangPRB2018, wuSci2016,xuPRA2016_Dirac,xuPRA2016_typeII}. One pair (the other) contains a laser beam with the 
Rabi frequency proportional to $\sin(k_R x)$ [$\cos(k_R x)$]
and another beam proportional to $e^{i k_R y}$. Given that our system is effectively 1D due to a
strong confinement along $y$ and $z$, 
the effects of the phase $e^{i k_R y}$ can be negligible. 
The Hamiltonian $H_{\textrm{c} }^{\textrm{lin} }(x,\theta)$ is invariant 
under translation by $2a$ with $a=\pi/k_R$. Despite this, we show that its Bloch states
can still be characterized by the momentum $k$ within the interval $[0,2\pi/a]$,
due to the symmetry
represented by $T_a \sigma_z$, where $T_a$ is the translation operator by $a$
(see Supplementary Note 6).
As a result, for the linear Thouless pumping in this model,
the average displacement per particle 
over one pump period
is $x_c=C a$, where $C$ is the Chern number for the corresponding Bloch band
(see Supplementary Note 6).

\begin{figure}
	\includegraphics[width=1\linewidth]{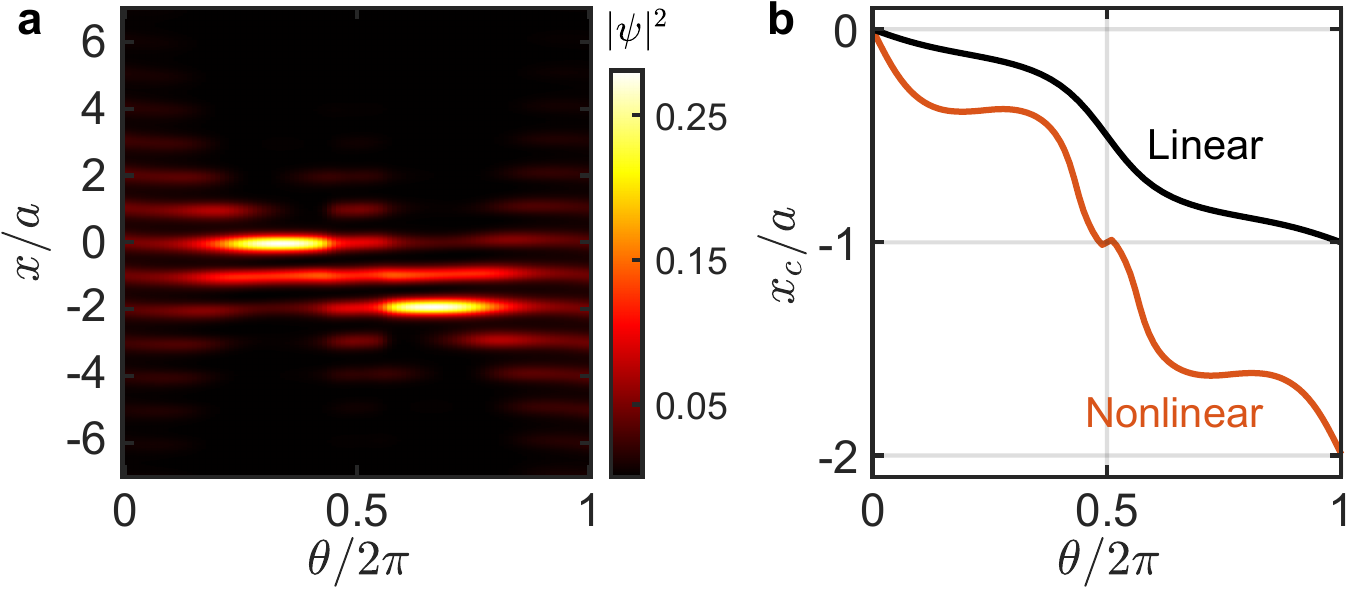}
	\caption{\textbf{Anomalous nonlinear pumping in the continuous model.} \textbf{a,} 
		One-cycle evolution of the density distribution $|\psi(x)|^2=\sum_{\sigma}|\psi_\sigma(x)|^2$ 
		of the instantaneous solitons bifurcating from the second band when $g_{12}=0$. 
		Note that the anomalous nonlinear pumping can still occur when $g_{12}\neq 0$.
		\textbf{b,} The evolution of center-of-mass positions $x_c$ of the instantaneous solitons (red line)
		and the corresponding Wannier functions (black line) with respect to $\theta$ over one period.
		Here, ${N}=0.2$ and $g=-1$.}
	\label{fign6}
\end{figure}

The continuous model can be mapped to a tight-binding model based on the 
Wannier functions $\{W_{nj}(x):j\in\mathbb{Z}\}$ from 
the lowest Bloch band of the Hamiltonian $H_0(x)$
with the band index $n=1$. This yields precisely the Hamiltonian in Eq.~(\ref{HLin}) 
with the parameters 
$J_1=-2 \int {\rm d} xW_{10}(x) H_0(x)W_{11}(x)$, $J_1^\prime =-2 V_s \int {\rm d} x W_{10}(x) \sin(k_Rx) W_{11}(x)$, 
$J_2= V_c \sin \theta \int {\rm d} xW_{10}(x) \cos(k_Rx)W_{10}(x)$, and $m_z=h_z$
(see the derivation in Supplementary Note 6).
Thus, we expect the emergence of anomalous nonlinear soliton pumping in the continuous model
in the presence of interactions described by nonlinear terms. 
In our study, we set $V_x=4 E_R$, $V_s=1.04 E_R$ and $V_c=0.21 E_R$, where $E_R=\hbar^2 k_R^2/(2m)$ 
is the recoil energy. The parameter $h_z$ is varied as $h_z=2t_x(1+\cos \theta )$ with $t_x=0.0855 E_R$.
Under these conditions, we find that the Chern number of the first and second bands of the continuous 
model are $C=1$ and $C=-1$, respectively. As a result, the average displacement for the linear Thouless pumping over
one cycle is
$a$ and $-a$, respectively (see Fig.~\ref{fign6}b).

We now demonstrate the emergence of anomalous nonlinear pumping in the 
continuous model in the presence of nonlinearity. The dynamics is governed by the following 
dimensionless Gross-Pitaevskii (GP) equation:
\begin{eqnarray}
	i \frac{\partial}{\partial t}\psi_\sigma (x)= 
	[\tilde{H}_{\textrm{c} }^{\textrm{lin} }\psi (x)]_\sigma 
	+g|\psi_{\sigma}(x)|^2\psi_{\sigma}(x),
\end{eqnarray}
where 
$\psi=\left(\psi_1(x),\psi_2(x)\right)^T$ and $\tilde{H}_{\textrm{c} }^{\textrm{lin} }$ is the dimensionless 
version of ${H}_{\textrm{c} }^{\textrm{lin} }$. 
The units of energy, time and length are $E_R$, $\hbar/E_R$ and $a$, respectively.
We have taken $g_{11}=g_{22}=g$ for simplicity and 
have confirmed that the results remain valid when they are slightly different.
We set $g=1$ and $g=-1$ for repulsive and attractive interactions, respectively.
The strength of nonlinearity is characterized by the norm $N=\sum_{\sigma}\int {\rm d}x|\psi_{\sigma}(x)|^2$.
For instance, consider a $^7\text{Li}$ BEC containing $1316$ atoms with
the scattering length $a_s \approx -1.43$ nm
under optical lattices along $x$ with $a=532$ nm and 
a transverse harmonic trap along $y$ and $z$ with the frequency $\omega_\perp=2\pi \times 710$ Hz~\cite{streckerNat2002,khaykovichSci2002}.
In this case, we have $N=0.2$.

By solving the nonlinear eigenstates of 
the instantaneous nonlinear Hamiltonian, we find that when $g>0$, the displacement of a soliton 
bifurcating from the first band over one period is $a$, consistent with linear Thouless
pumping. However, the corresponding tight-binding model produces a nonlinear pumping displacement
of $2a$, suggesting that the tight-binding model may oversimplify the nonlinear effects 
(see the discussion in Supplementary Note 6). 
Remarkably, we find that when $g<0$, a soliton that bifurcates from the second band in the continuous 
model is pumped across $-2a$ over one cycle, as depicted in Fig.~\ref{fign6}, which is 
twice the Chern number of the second band. 
We also find that for negative $g$, a soliton coming from the first band 
(whose Chern number is $1$)
does not exhibit any pumping over one cycle (see Supplementary Note 1),
corresponding to the first anomalous case shown in Fig.~\ref{fign1}a and Fig.~\ref{fign2}b.
Furthermore, 
we have numerically verified 
the stability of the instantaneous soliton solutions as $\theta$ runs from $0$ to $2\pi$. 
In addition, the numerical time evolution of an initial soliton is also found to be in 
agreement with the instantaneous solution. 
These results suggest that anomalous nonlinear pumping can be experimentally 
observed in the continuous model. 

In summary, we have demonstrated the emergence of anomalous nonlinear soliton pumping, wherein the 
correspondence between the quantized displacement of a soliton and the Chern number of 
the underlying linear Bloch band breaks down. 
Specifically, a soliton can be pumped across one, two or three unit cells over one pumping cycle, 
while the underlying linear band carries the Chern number of $-1$.
This anomalous behavior is beyond the previous understanding of nonlinear 
pumping of a soliton based on the center-of-mass flow of the instantaneous Wannier functions. 
We show that the anomalous nonlinear pumping occurs due to 
the transition of a soliton between different Wannier functions by passing through an intersite-soliton state.
Furthermore, we illustrate that
nonlinearity can induce the quantized pumping of a soliton, even when the corresponding 
linear Bloch band is topologically trivial. 
We note that our results are completely different from the phenomenon in a Bose-Bose atomic mixture in Ref.~\cite{mostaanNC2022} 
where a soliton of impurity atoms can undergo a 
nonzero quantized displacement during a pump cycle, as a result of the dragging from a soliton of majority atoms that bifurcates 
from a topologically nontrivial band. There, the pumping can also be understood as 
the center-of-mass flow of the Wannier functions of the corresponding Bloch band of the linear Hamiltonian
felt by the majority atoms. 
In our case, we have only one set of band structures, and the discrepancy between the band topology and 
quantized motion emerges spontaneously, going beyond the description of the Wannier function of
the original linear Hamiltonian.  

Our work opens a new avenue for studying nonlinear pumping and inspires a range 
of intriguing directions for future research. From an experimental perspective, 
given that anomalous nonlinear pumping can occur in both discrete and continuous models, 
it can be observed in either photonic systems or ultracold atomic gases. 
Theoretically, a broad class of interesting topics await further exploration, 
including multi-component nonlinear pumping beyond the two-component setting, 
anomalous fractional soliton pumping, and anomalous nonlinear pumping in higher 
dimensions, and general principles that guarantee the stability of intersite solitons in 
two-component systems.

\textbf{Data Availability:} The data that support the findings of this study are available
at Figshare \url{https://doi.org/10.6084/m9.figshare.32182566} (ref.~\cite{ref-data}).


%

\textbf{Acknowledgments:}
We thank Yongping Zhang and Yan-Bin Yang for helpful discussions.
We also acknowledge the support by center of high performance computing, Tsinghua University.

\textbf{Funding Statements:} 
This work is supported by 
Quantum Science and Technology-National Science and Technology Major Project (Grant No. 2021ZD0301604)
and the National Natural Science Foundation of China (Grant No. 11974201).

\textbf{Author Contributions:} 
Y.L.T., J.H.W., and Y.X. contributed to all aspects of this work. Y. Xu initiated and supervised the project.

\textbf{Competing Interests:} The authors declare that there are no competing interests.

\textbf{Author Information:} 
Correspondence and requests for materials should be addressed to Y.X. (yongxuphy@tsinghua.edu.cn).

\begin{widetext}

\section{Supplementary Information}

\renewcommand{\figurename}{{Supplementary Figure}}
\renewcommand{\tablename}{Supplementary Table}

\setcounter{equation}{0} \setcounter{figure}{0} \setcounter{table}{0} %
\renewcommand{\theequation}{S\arabic{equation}}
\renewcommand{\thefigure}{%
	\arabic{figure}}
\renewcommand{\bibnumfmt}[1]{[#1]}
\renewcommand{\citenumfont}[1]{#1}

\section{Supplementary Note 1. Comparison with the trapped soliton for strong nonlinearity}
In the main text, we have presented a trapped-like nonlinear soliton pumping where a displacement
of a soliton vanishes over one cycle, even though the underlying linear band is topologically nontrivial,
and nonlinearity is not strong. This can occur for a soliton coming from the lowest band 
of the discrete model and the continuous model when $g<0$ and $g_{12} = 0$.
We have also shown that it occurs due to the transition to an intersite soliton. 
The phenomenon is reminiscent of the trapped case
occurring for strong nonlinearity~\cite{jurgensenNat2021, jurgensenNP2023, fuPRL2022}. 
However, the latter scenario occurs because it occupies all Bloch bands,
resulting in a total Chern number of zero (see Fig.~2d in Ref.~\cite{jurgensenNP2023}). 
In contrast, in our case, a soliton mainly occupies a single band, and thus  
the mechanism is completely different.

\begin{figure}[htp]
	\includegraphics[width = \linewidth]{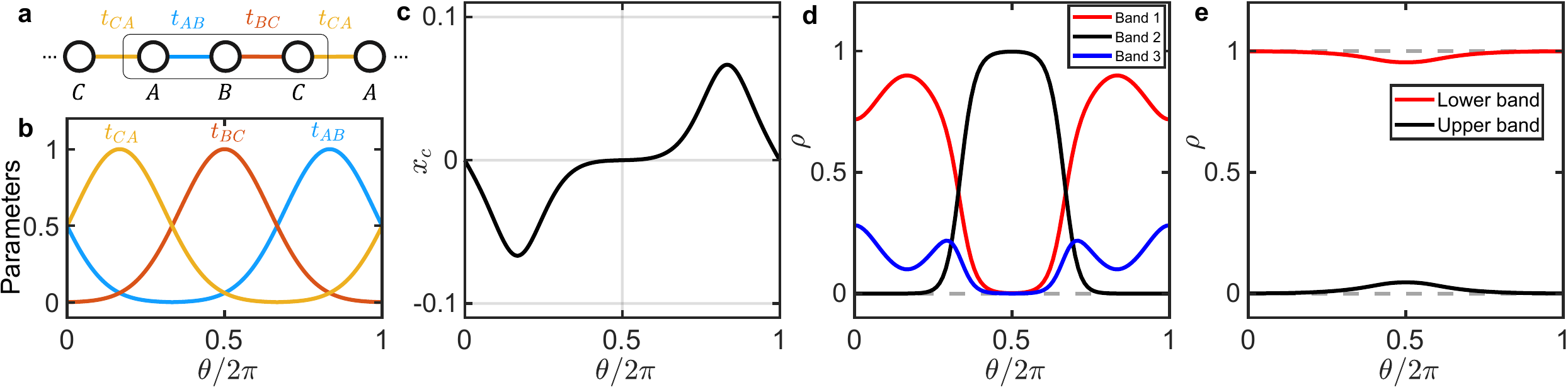}
	\caption{\textbf{a,} Schematic illustration of the off-diagonal AAH model. \textbf{b,} Intra-cell and 
		inter-cell couplings in the off-diagonal AAH model as a function of $\theta$ from $0$ to $2\pi$. 
		\textbf{c,} The center-of mass 
		trajectory of instantaneous solitons bifurcating from band 1 with $N=2.5$ in the off-diagonal AAH model. 
		\textbf{d,} Occupations of instantaneous solitons on band $i$ with $i=1,\ 2$, and 3 as a function of $\theta$ 
		corresponding to the case in 
		\textbf{c}. \textbf{e,} Occupations of instantaneous solitons on the lower and upper bands 
		versus $\theta$ corresponding to case 1 in Fig.~2 in the main text.}
	\label{Fig_AAH}
\end{figure}

For better readability, we follow Refs.~\cite{jurgensenNat2021, jurgensenNP2023, fuPRL2022} to 
demonstrate that the trapped pumping found in Ref.~\cite{jurgensenNat2021} 
is attributable to occupation of a soliton on all the linear bands. 
Specifically, we consider the nonlinear off-diagonal Aubry-Andr\'{e}-Harper (AAH) model~\cite{jurgensenNat2021},
which is a 1D tight-binding model whose unit cell consists of three sites 
labelled by $A$, $B$, and $C$ (see Supplementary Fig.~\ref{Fig_AAH}a). The sites are connected by the intra-cell coupling 
$t_{AB}(\theta)$ and $t_{BC}(\theta)$ and
inter-cell coupling $t_{CA}(\theta)$. These parameters are periodic functions of $\theta$ with 
a period of $2\pi$ which are displayed in Supplementary Fig.~\ref{Fig_AAH}b. Thus, the real-space linear Hamiltonian is expressed as
\begin{align}
	\label{AAH}
	{H}_{\textrm{AAH}}(\theta)=&\sum_{\jr}[t_{AB}(\theta)|\jr,B\rangle \langle \jr,A| +t_{BC}(\theta) |\jr,C \rangle \langle \jr,B|
	+\text{H.c.}]\\ \nonumber
	&+\sum_{\jr}[t_{CA}(\theta) |\jr+1,A\rangle \langle \jr,C|+\text{H.c.}].
\end{align}
This linear model has three nontrivial bands, and the corresponding  
Chern numbers, denoted as $C_i$ with $i=1,\ 2$, and 3, exhibit the values: $C_1=C_3=-1$ and $C_2=2$.

In the presence of nonlinearity, the time-dependent nonlinear Schr\"{o}dinger equation is given by
\begin{align}
	\label{NSL}
	i \frac{\partial}{\partial t} \phi_{\sigma \jr}(t)=\sum_{\jrp,\sigma'}[H_{\textrm{AAH}}(\theta)]_{\sigma \jr,\sigma' \jrp}\phi_{\sigma' \jrp}(t)+g|\phi_{\sigma \jr}(t)|^2\phi_{\sigma \jr}(t),
\end{align}
where $\phi_{\sigma \jr}(t)$ is the value of a wavefunction at site $\sigma$ of unit cell $\jr$ at time $t$. Here, 
we consider the focusing nonlinearity with $g=-1$ and use the norm $N=\sum_{\sigma \jr}|\phi_{\sigma \jr}|^2$ 
to describe the strength of nonlinearity.

For the strongly nonlinear case with $N=2.5$, we plot the center-of-mass trajectory
of instantaneous solitons bifurcating from linear band 1 in Supplementary Fig.~\ref{Fig_AAH}c.
We clearly see that the soliton returns to the starting position at
the end of the pumping. 
We now compute the band occupation of instantaneous soliton solutions $\phi(\theta)$ on the linear 
band $i$ with $i=1,\ 2$, and 3,
\begin{align}
	\label{P_i}
	\rho_i(\theta)=\frac{1}{N}\sum_{k}| \sum_{\sigma \jr} \varphi_{i,k,\sigma,\jr}(\theta)^*\phi_{\sigma \jr}(\theta)|^2,
\end{align}
where $\varphi_{i,k,\sigma,\jr}(\theta)$ is the $i$th band's Bloch state at momentum $k$. 
The results are presented in Supplementary Fig.~\ref{Fig_AAH}d, illustrating that 
all linear bands are involved for the soliton over one cycle.
If we assume that the soliton follows the multiband Wannier function, 
then it does not exhibit any displacement over a cycle since the total Chern
number of all bands vanishes~\cite{jurgensenNP2023}.

However, in our case, the lower band dominates the occupation in the 
entire pumping period as shown in Supplementary Fig.~\ref{Fig_AAH}e. This indicates that 
the argument based on
multiband occupations is not applicable to the trapped behavior observed in our system. 
In fact, it belongs to the anomalous nonlinear pumping, which has been elucidated 
in the main text. Furthermore, we consider the continuous model in Eq.~(5) in the main text, 
and the anomalous nonlinear pumping where the soliton comes from the first band 
with negative $g$ ($g=-1$) is presented in Supplementary Fig.~\ref{Fig_conti_trap}. 
Same as the case in the tight-binding model, the soliton is dominated by 
the first band with +1 Chern number throughout the pumping period, 
while its displacement over one cycle is zero.

\begin{figure}[htp]
	\includegraphics[width =0.8 \linewidth]{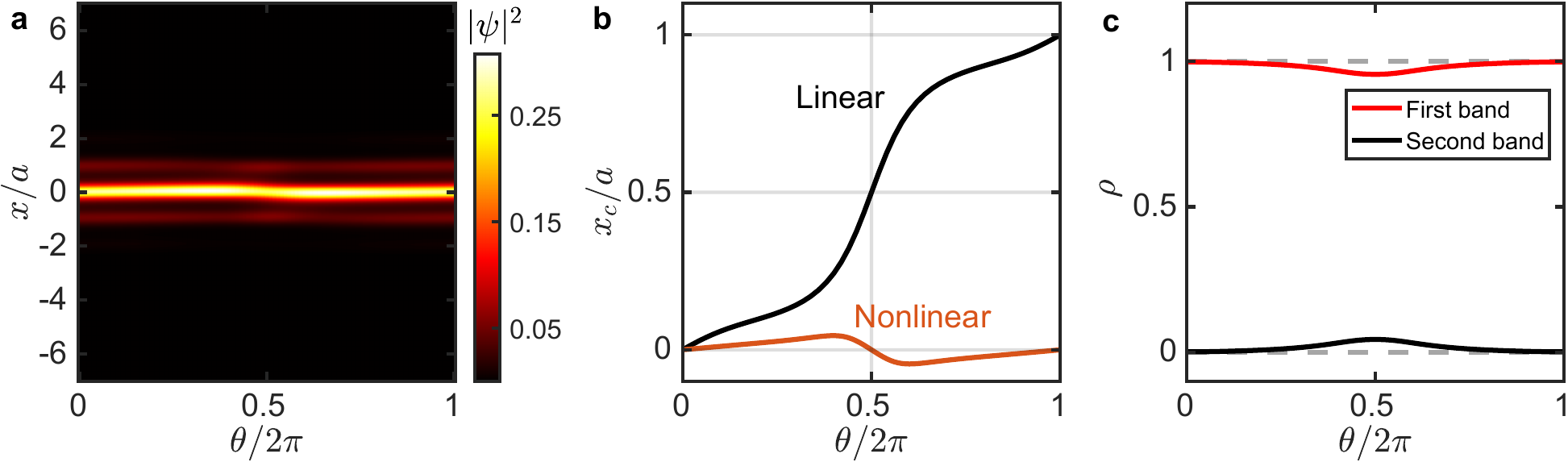}
	\caption{
		\textbf{a,} One-cycle evolution 
		of the density distribution $|\psi(x)|^2=\sum_{\sigma}|\psi_\sigma(x)|^2$ 
		of the instantaneous solitons 
		bifurcating from the first band in the continuous model in Eq.~(4) in the main text. 
		\textbf{b,}	The evolution of center-of-mass positions $x_c$ of the instantaneous solitons (red line) and the corresponding Wannier
		functions (black line) with respect to $\theta$ over one period.
		\textbf{c,} Occupations of instantaneous solitons on the first and second bands	
		versus $\theta$ corresponding to the case in \textbf{a} and \textbf{b}.
		Here, we set $g=-1$, $g_{12}=0$, and $N=0.2$.
	}
	\label{Fig_conti_trap}
\end{figure}

\section{Supplementary Note 2. Stability analysis}
In this section, we will follow Ref.~\cite{kevrekidisBook2009} to perform the stability analysis for the soliton solutions at a fixed $\theta$,  
$\psi^{(0)}(t)=e^{-i \mu t} \chi^{(0)}(\theta)$ where $\mu$ is the chemical potential
and $\chi^{(0)}$ satisfies the following stationary nonlinear equation:
\begin{equation}
	\mu \chi_{\sigma \jr}^{(0)} = 
	\sum_{\sigma^\prime \jrp } 
	H_{\sigma \jr, \sigma^\prime \jrp }^{\textrm{lin}}(\theta) \chi_{\sigma^\prime \jrp}^{(0)}
	+\left(g|\chi_{\sigma \jr}^{(0)}|^2+g_{12} |\chi_{\overline{\sigma} \jr}^{(0)}|^2\right)\chi_{\sigma \jr}^{(0)}.
\end{equation}
We now evolve the state with a small perturbation, $\psi(t) =e^{-i \mu t} (\chi^{(0)} + \delta \psi(t))$, yielding
\begin{equation} \label{Pert-Eq}
	i\partial_t \delta \psi_{\sigma \jr}(t)  =[H^\prime \delta \psi ]_{\sigma \jr} + 
	\left[H_1 \left(\begin{array}{c}
		\delta \psi_{1\jr}\\
		\delta \psi_{2\jr} \\
	\end{array}\right)+H_2 
	\left(\begin{array}{c}
		\delta \psi_{1\jr}^*\\
		\delta \psi_{2\jr}^* \\
	\end{array}\right) \right]_\sigma, 
\end{equation}
where $H^\prime = H^\textrm{lin}-\mu$ and
\begin{eqnarray}
	H_1&=&\left(\begin{array}{cc}
		2g{|\chi_{1 \jr}^{(0)}|}^2 +g_{12}{ |\chi_{2 \jr}^{(0)}| }^2 & g_{12} \chi_{1 \jr}^{(0)} \chi_{2 \jr}^{(0)*}\\
		g_{12} \chi_{1 \jr}^{(0)*} \chi_{2 \jr}^{(0)} & 2g|\chi_{2 \jr}^{(0)}|^2 +g_{12} |\chi_{1 \jr}^{(0)}|^2 \\
	\end{array}\right), \\
	H_2&=&\left(\begin{array}{cc}
		g(\chi_{1 \jr}^{(0)})^2  & g_{12} \chi_{1 \jr}^{(0)} \chi_{2 \jr}^{(0)}\\
		g_{12} \chi_{1 \jr}^{(0)} \chi_{2 \jr}^{(0)} & g (\chi_{2 \jr}^{(0)})^2 \\
	\end{array}\right).
\end{eqnarray}
To evaluate the spectrum, we write $\delta \psi_{\sigma \jr}(t)$ as
$
\delta \psi_{\sigma \jr}(t)=u_{\sigma \jr} e^{-i\omega t }+v_{\sigma \jr}^* e^{i \omega^* t}
$
with $\omega$ being the excitation frequency
and substitute it into Eq.~(\ref{Pert-Eq}) to arrive at the following Bogoliubov-de Gennes (BdG) equation
\begin{equation} \label{Eq:BdG}
	\begin{split}
		\omega  u_{\sigma \jr} = [H^\prime u]_{\sigma \jr} + [H_1 u_\jr]_{\sigma} +[H_2 v_\jr]_\sigma, \\
		\omega  v_{\sigma \jr} = -[H^{\prime *} v]_{\sigma \jr} -[H_1^* v_\jr]_{\sigma} -[H_2^* u_\jr]_\sigma.
	\end{split}
\end{equation}
Solving the equation gives us the excitation spectrum. When the system possesses excitations with complex 
energy, the system is unstable. 
We have numerically calculated the excitation spectrum for instantaneous 
soliton solutions for anomalous nonlinear pumping as the system parameter $\theta$ is varied from
$0$ to $2\pi$. We find that the maximum $\omega_m=\text{max}(|\text{Im}(\omega)|)$
of the absolute value of the imaginary part of $\omega$ is smaller than
$10^{-7}$ in the entire region of $\theta$ from $0$ to $2\pi$, indicating that the instantaneous soliton solutions are
stable.

\section{Supplementary Note 3. Occupations of instantaneous solitons on Wannier functions}
In Fig.~2 in the main text, we have presented the occupations of instantaneous solitons on 
Wannier functions along the Wannier centers. To see this more clearly, we provide
the occupations of the solitons on three neighboring Wannier functions as a function of $\theta$
in Supplementary Fig.~\ref{fig_occ_Wann} for the four cases in Fig.~2 in the main text.
One can clearly see that the soliton transitions to an ideal intersite soliton at $\theta=\pi$ 
in Supplementary Figs.~\ref{fig_occ_Wann}b and c.

\begin{figure}[htp]
	\includegraphics[width = \linewidth]{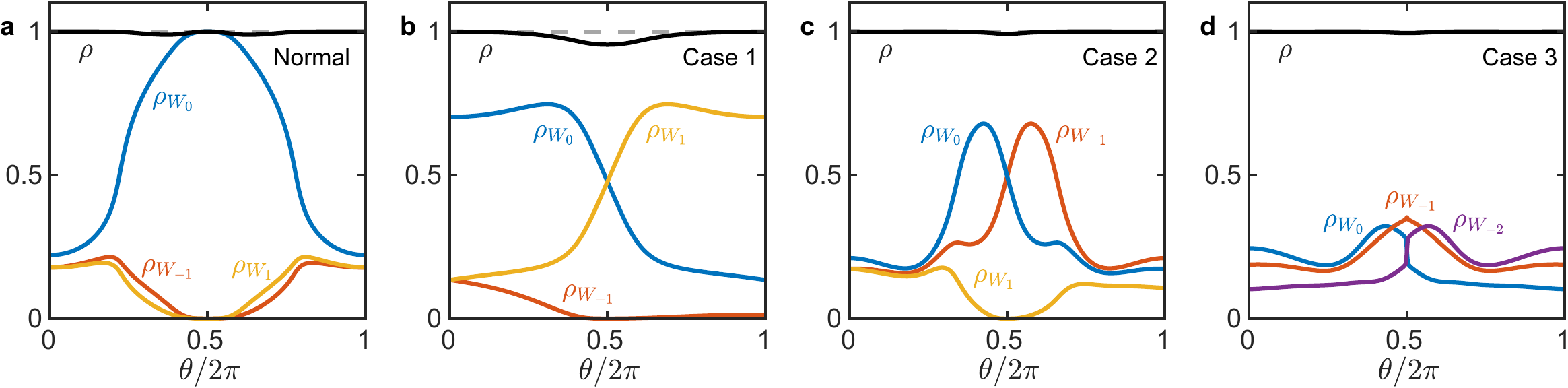}
	\caption{
		Occupations $\rho_{W_l}(\theta)$ of instantaneous solitons on 
		three Wannier functions of the lowest band of the linear Hamiltonian in Eq. (2) in the main text
		with respect to $\theta$ for the four cases. The parameters are the same as those in
		Fig. 2 in the main text. The black lines represent the occupation on the lowest band,
		showing that the solitons primarily occupy the lowest band.
	}
	\label{fig_occ_Wann}
\end{figure}

\section{Supplementary Note 4. Effects of $g$ and a phase diagram}
\subsection{A. Does a soliton exist when $g=0$?}
In the main text concerning the discrete nonlinear model, we set $g=1$. We find that for the soliton solution 
$(\psi_{1j},\psi_{2j})^T$ , $|\psi_{2j}|^2$ is much larger than 
$|\psi_{1j}|^2$ when $\theta=0$, as shown in Supplementary Fig.~\ref{fig_otherg}. Consequently, the nonlinearity 
for the second component in the nonlinear equation is 
approximated by $g|\psi_{2j} |^2$. Hence, $g$ plays a key 
role in determining the soliton's width. As $g$ decreases, 
$|\psi_{2j} |^2$ broadens (see Supplementary Fig.~\ref{fig_otherg}), 
suggesting that no soliton solution exists in this nonlinear equation when $g=0$ at $\theta=0$.

\begin{figure}[htp]
	\includegraphics[width = 0.3 \linewidth]{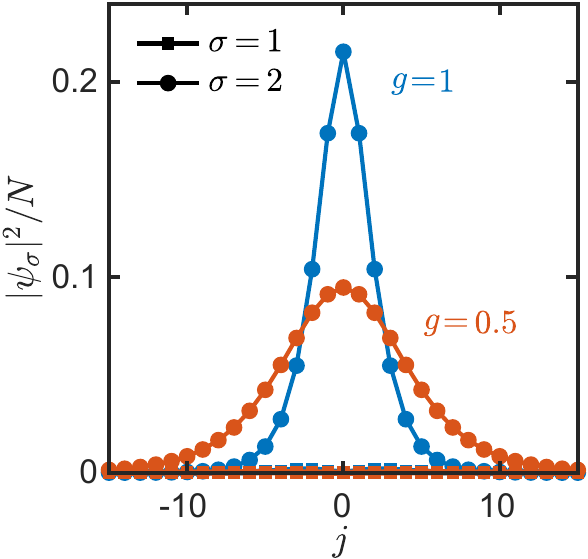}
	\caption{
		The density distribution of the solitons for different components 
		at $\theta=0$ for $g=1$ (blue line) and $g=0.5$ (red line). 
		Here, we set $m_0=g_{12}=1$ and $N=1.45$.
	}
	\label{fig_otherg}
\end{figure}	

\subsection{B. A phase diagram}
We provide a phase diagram of the anomalous nonlinear soliton pumping 
with respect to $m_0$ and $g_{12}$ in Supplementary Fig.~\ref{fig_phasediagram}.

\begin{figure}[htp]
	\includegraphics[width = 0.45 \linewidth]{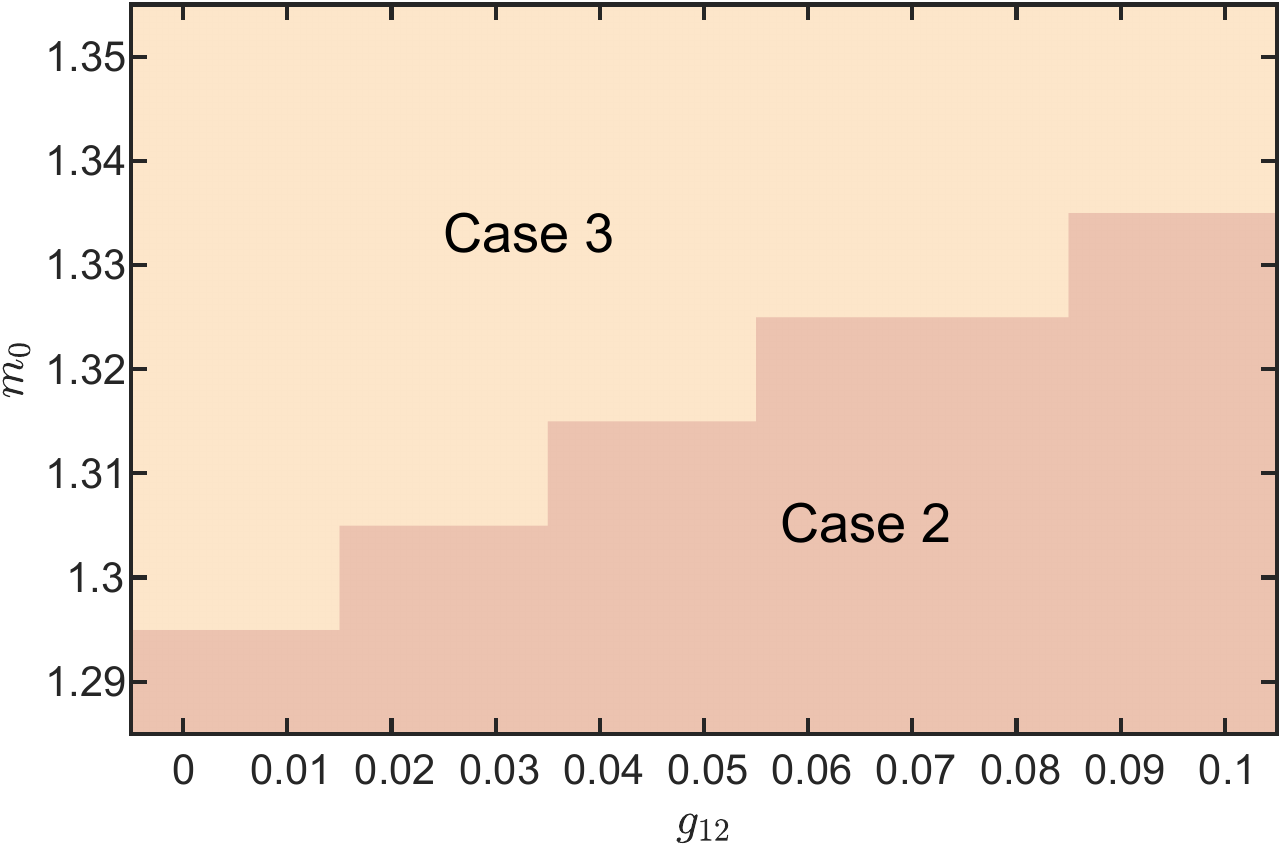}
	\caption{The phase diagram of the anomalous nonlinear soliton pumping 
		with respect to $m_0$ and $g_{12}$.
		Here, we set $g=1$ and $N=1.45$.
	}
	\label{fig_phasediagram}
\end{figure}

\section{Supplementary Note 5. The intersite soliton} 
\label{Sec-SSH-non}

\subsection{A. Derivation of the intersite soliton}
In the main text, we state that when $\theta=\pi$, there are soliton solutions localized near $j=l+1/2$, 
which are identical to a Wannier function, and solutions localized near $j=l$, where $l$ is an integer. 
In this section, to elucidate
the statement, we transform the linear Hamiltonian $H^{\textrm{lin}}$ to the SSH model at this
$\theta$, that is, $H^{\textrm{lin}}_{\textrm{SSH}}=UH^{\textrm{lin}}U^\dagger$
with $U=e^{-i\pi\sigma_y/4}$. In momentum space, we have $H^{\textrm{lin}}_{\textrm{SSH}}(k)=(m_z+\cos k)\sigma_x+\sin k \sigma_y$. 
Accordingly, the time-dependent nonlinear Schr\"odinger equation is 
transformed to 
\begin{equation} \label{SSH-Eq}
	i\frac{\partial}{\partial t}\phi_{\sigma \jr}= 
	[H_{\textrm{SSH} }^{\textrm {lin}} \phi ]_{\sigma \jr}
	+\left(g^{\textrm{SSH}} |\phi_{\sigma \jr}|^2+g_{12}^{\textrm{SSH} } |\phi_{\overline{\sigma} \jr}|^2\right) \phi_{\sigma \jr},
\end{equation}
where $\phi_\jr=U\psi_\jr$ with $\phi_\jr=\left(\phi_{1\jr},\phi_{2\jr}\right)^T$ and $\psi_\jr=\left(\psi_{1\jr},\psi_{2\jr}\right)^T$, 
$g^{\textrm{SSH}}=(g+g_{12})/2$ and $g_{12}^{\textrm{SSH} }=(3g-g_{12})/2$. Since 
$H_{\textrm{SSH} }^{\textrm {lin}}$ in real space is a real matrix, we are only interested in the solutions that are real. 
Supplementary Fig.~\ref{figs2}a illustrates the SSH model with each unit cell containing two sites labeled by $1$ and $2$.
When $m_0=1$ and thus $m_z=0$, the model in real space contains 
only the intercell hopping as shown in Supplementary Fig.~\ref{figs2}b. 
In this case, the Wannier functions are localized around  $j=l+1/2$ between two adjacent unit cells. 
Their wavefunctions are
$W_{\lr}^{(\textrm{l} )}=(|2,\lr\rangle -|1,\lr+1\rangle )/\sqrt{2}$ and 
$W_{\lr}^{( \textrm{u} )}=(|2,\lr\rangle+|1,\lr+1\rangle)/\sqrt{2}$ for the lower and upper bands, respectively.
Here $|\sigma,\lr\rangle$ with $\sigma=1, 2$ describes the degree of freedom at the $\sigma$th site 
in the $\lr$th unit cell.
The Wannier functions corresponding to $\phi_{2,\lr}=1/\sqrt{2}$ and $\phi_{1,\lr+1}=\mp 1 /\sqrt{2}$ (all other entries in $\phi$ are equal to zero) 
are clearly the solutions of the stationary nonlinear equation: 
\begin{eqnarray}
	\mu \phi_{2,\lr}=\phi_{1,\lr+1}+(g^{\textrm{SSH} } \phi_{2,\lr}^2+ g_{12}^{\textrm{SSH} } \phi_{1, \lr}^2) \phi_{2, \lr}, \\
	\mu \phi_{1,\lr+1}=\phi_{2,\lr}+(g^{\textrm{SSH} } \phi_{1,\lr+1}^2+g_{12}^{\textrm{SSH} } \phi_{2, \lr+1}^2) \phi_{1,\lr+1},
\end{eqnarray}	
where $\mu=(g^{\textrm{SSH} }/2) \mp 1$ is the instantaneous eigenvalue.
If we require that $N=\sum_{\sigma \jr} |\phi_{\sigma \jr}|^2$, then $\sqrt{N} W_{\lr}^{(\textrm{l})}$
and $\sqrt{N} W_{\lr}^{(\textrm{u})}$ are the solutions to this 
nonlinear equation with $\mu=(Ng^{\textrm{SSH} }/2) - 1$ and $\mu=(Ng^{\textrm{SSH} }/2) + 1$,
respectively. However, this solution becomes unstable when $g_{12}$ becomes small. 

In the main text, we have demonstrated the emergence of a new branch of stable soliton solutions localized
between two neighboring unit cells when $\theta=\pi$. In the following, we will illustrate that these soliton solutions 
are mainly a superposition of two adjacent Wannier functions. 
To obtain the new nonlinear solution localized around $j=l$, we expand $\phi$ using adjacent lower-band and upper-band Wannier functions 
$W_{\lr-1}^{(\textrm{l},\textrm{u})}$ and $W_{\lr}^{(\textrm{l},\textrm{u})}$ as
\begin{equation}
	\label{wul_1}
	\phi=c_\textrm{l} W_\lr^{(\textrm{l})}+c_\textrm{u} W_\lr^{(\textrm{u})}-c_\textrm{l} W_{\lr-1}^{(\textrm{l})}+c_\textrm{u} W_{\lr-1}^{(\textrm{u})},
\end{equation}
where we have assumed that the solution is symmetric about $x=n$ so that there are only two independent variables $c_\textrm{l}$ and $c_\textrm{u}$.
Since $\sum_{\sigma \jr}| \phi_{\sigma \jr} |^2=N$, we can write 
\begin{eqnarray}
	\label{wul2}
	c_\textrm{l} &=&\frac{\sqrt{N}}{2}\left(\sin\varphi+\cos\varphi\right),\\ 
	c_\textrm{u} &=& \frac{\sqrt{N}}{2}\left(\sin\varphi-\cos\varphi\right).
\end{eqnarray}
$e^{-i\mu t}\phi$ is a stationary solution of Eq.~(\ref{SSH-Eq}) if $\varphi$ satisfies the following transcendental equation:
\begin{align}
	\label{varphi2}
	\frac{(3g-g_{12})N}{2}+\frac{(5g+g_{12})N}{2}\cos2\varphi+8\cot2\varphi=0.
\end{align}

\begin{figure}[htp]
	\includegraphics[width = 0.6\linewidth]{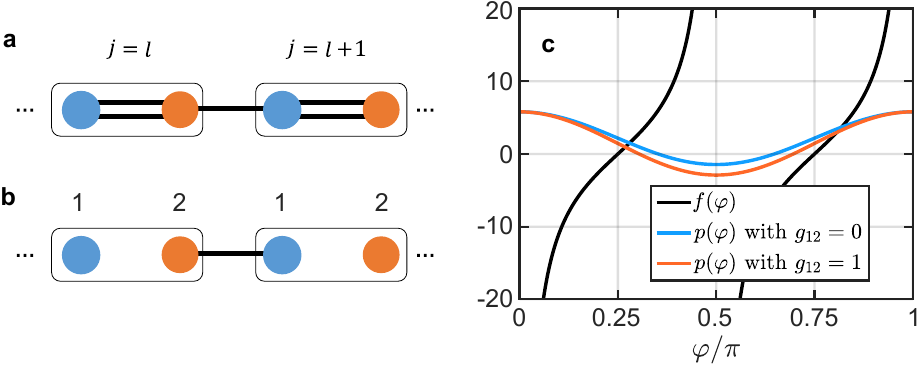}
	\caption{\textbf{a,b,} Schematic illustration of the SSH model. Each unit cell consists of two sites labeled
		by $1$ and $2$.
		\textbf{c,} Plot of $f(\varphi)$ and $p(\varphi)$ as a function of $\varphi$ with their intersections representing
		the solutions of the transcendental equation (\ref{varphi2}). Here, we set $N=1.45$ and $g=1$.}
	\label{figs2}
\end{figure}

To illustrate the particular solution $\varphi$ of Eq.~(\ref{varphi2}), we define two functions of $\varphi$,
\begin{eqnarray}
	\label{fpvarphi}
	f(\varphi)&=&-8\cot2\varphi,\\
	p(\varphi)&=&\frac{(3g-g_{12})N}{2}+\frac{(5g+g_{12})N}{2}\cos2\varphi,
\end{eqnarray}
which are plotted in Supplementary Fig.~\ref{figs2}c by taking $N=1.45$. 
We see that they intersect around $\varphi=n\pi+\pi/4$ and $n\pi+3\pi/4$, representing 
particular solutions of Eq.~(\ref{varphi2}). 
The former (latter) ones correspond to solutions dominated by lower (upper) band Wannier functions. 
For example,
when $g_{12}=0$ and $N=1.45$, we have $\varphi=1.118\times\pi/4$, and thus 
$c_\textrm{l}=0.848$, and $c_\textrm{u}=0.079$.
The figure also indicates that the solutions change very slightly when $g_{12}$ is varied from $0$ to $1$.
However, we find that the solution is stable only when $g_{12}$ is small.

For the trapped-like case, since numerical results suggest that the soliton solution is antisymmetric 
about $j=l$, we write the solution in terms of Wannier functions as 
\begin{equation}
	\label{wul_t}
	\phi=-c_\textrm{l}W_\lr^{ \textrm{(l)} }+c_\textrm{u} W_\lr^{\textrm{(u)}}
	-c_\textrm{l}W_{\lr-1}^{\textrm{(l)}}-c_\textrm{u} W_{\lr-1}^{\textrm{(u)}}.
\end{equation}
Similarly, we derive a transcendental equation that $\varphi$ should satisfy:
\begin{align}
	\label{varphi3}
	\frac{(g_{12}-3g)N}{2}+\frac{(5g+g_{12})N}{2}\cos2\varphi+8\cot2\varphi=0.
\end{align}
This equation also exhibits solutions near $n\pi+\pi/4$, corresponding to solitons dominated by 
lower-band Wannier functions. For example, when $g=-1$, $g_{12}=0$ and $N=1.45$, we find that 
$\varphi=1.275\times\pi/4$ so that $c_\textrm{l}=0.832$, and $c_\textrm{u}=0.183$. Similar to the previous
case, the soliton solutions are stable only when $|g_{12}|$ is small.

\subsection{B. Stability analysis of onsite and intersite solitons}
In this subsection, we will perform stability analysis 
for onsite and intersite solitons at $\theta=\pi$.
We first plot numerically calculated $\omega_m$ with respect to $g_{12}$ 
in Supplementary Fig.~\ref{Fig_omega_m}, illustrating that the on-site (intersite) soliton is stable 
with $\omega_m=0$ when $g_{12}>0.09$ ($g_{12}<0.24$).

\begin{figure}[htp]
	\includegraphics[width = 0.55\linewidth]{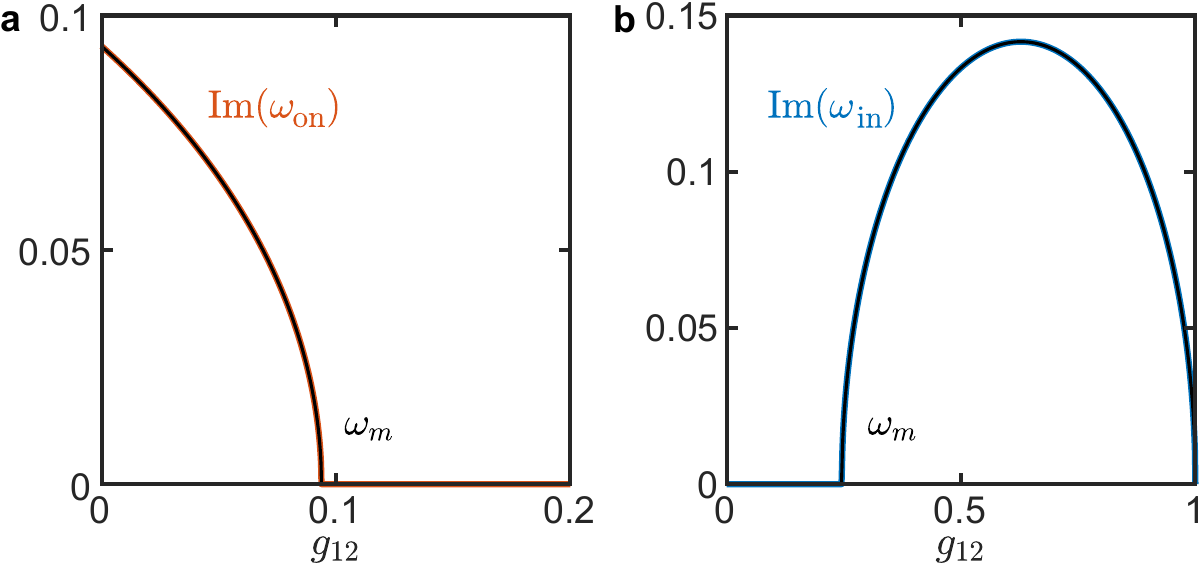}
	\caption{
		The maximum $\omega_m$ 
		of the absolute value of the imaginary part of the excitation frequency 
		for on-site (\textbf{a}) and intersite (\textbf{b}) solitons at $\theta=\pi$ with respect to $g_{12}$.
		The black lines are obtained by diagonalization of the full BdG Hamiltonian, while
		the red and blue lines are obtained by diagonalization of the restricted matrices in Eq.~(\ref{BdGsingle}) and 
		Eq.~(\ref{BdGOP}), respectively.
		Here, we set $m_0=g=1$ and $N=1.45$.
	}
	\label{Fig_omega_m}
\end{figure}

To derive the analytical expression of $\omega_m$ for the 
onsite and intersite solitons, 
we write the BdG Hamiltonian $\mathcal{L}$ in Eq.~(\ref{Eq:BdG}) in the basis of Wannier functions.
For the onsite soliton whose wavefunction is
$\phi_{\textrm{on}}=\sqrt{N}W_l^{(\text{l})}$,
we find that an unstable soliton has a pair of identical eigenvalues $\omega$
satisfying $\text{Re}(\omega)=0$ and $\text{Im}(\omega)>0$ and the 
corresponding eigenvectors have nonzero components only on 
$W_{l-1}^{(\textrm{l,u})}$ or $W_{l+1}^{(\textrm{l,u})}$. Thus, 
we restrict $\mathcal{L}$ to the subspace spanned by 
$\{ W_{l-1}^{(\textrm{l})},W_{l-1}^{(\textrm{u})}\}$ or $\{W_{l+1}^{(\textrm{l})}, W_{l+1}^{(\textrm{u})} \}$
so that the matrix $\mathcal{L}_{\textrm{on},l+1}$ and $\mathcal{L}_{\textrm{on},l-1}$ in the
corresponding subspace reads
\begin{align}
	\label{BdGsingle}
	\mathcal{L}_{\textrm{on},l\pm 1}=
	\begin{pmatrix}
		-g_b & \pm g_a & (g_a-g_{b})/2 & \pm(g_a-g_{b})/2 \\
		\pm g_a & 2-g_b & \pm(g_a-g_{b})/2 & (g_a-g_{b})/2 \\
		-(g_a-g_{b})/2 & \mp(g_a-g_{b})/2 & g_b & \mp g_a \\
		\mp(g_a-g_{b})/2 & -(g_a-g_{b})/2 & \mp g_a & g_b-2 \\
	\end{pmatrix}
\end{align}
with $g_a=gN/4$ and $g_b=g_{12}N/4$. They have the same characteristic equation,
\begin{align}
	\label{BdGsingleome}
	\omega^4+[4(g_b-1)-(g_a+g_b)^2]\omega^2+(g_a+g_b)(2g_a^2-2g_b^2-g_a+3g_b)=0.
\end{align}
We can analytically solve the eigenvalues $\pm\omega_{\textrm{on},i}$ with $i=1,2$, and the unstable excitations with imaginary $\omega_{\textrm{on}}$ emerge at $W_{l-1}$ and $W_{l+1}$. As shown in Supplementary Fig.~\ref{Fig_omega_m}a, the analytical result of $\text{Im}(\omega_{\textrm{on}})$ is in agreement with the numerical one.

For the intersite soliton $\phi_{\textrm{in}}$ composed of $W_{l-1}^{(\textrm{l,u})}$ and $W_{l}^{(\textrm{l,u})}$, the numerical result shows that its unstable excitation only consists of these Wannier functions. Thus, same as the on-site case, we restrict $\mathcal{L}$ to the subspace spanned by $W_{l-1}^{(\textrm{l,u})}$ and $W_{l}^{(\textrm{l,u})}$, and the restricted matrix is given by
\begin{align}
	\label{BdGOP}
	\mathcal{L}_{\textrm{in},l}=
	\begin{pmatrix}
		a-1 & b & c & b & d & e & f & e \\
		b & a-1 & -b & -c & e & d & -e & -f \\
		c & -b & a+1 & -b & f & -e & d & -e \\
		b & -c & -b & a+1 & e & -f & -e & d \\
		-d & -e & -f & -e & 1-a & -b & -c & -b \\
		-e & -d & e & f & -b & 1-a & b & c \\
		-f & e & -d & e & -c & b & -a-1 & b \\
		-e & f & e & -d & -b & c & b & -a-1
	\end{pmatrix}
\end{align}
with 
\begin{align}
	\label{BdGOPpara}
	\begin{split}
		a&=\frac{N}{4}(g+g_{12}+g\cos^2\varphi)-\frac{g^{\text{SSH}} N}{2}(\sin^4 \varphi+\cos^4 \varphi)-\frac{g^{\text{SSH}}_{12} N}{2}\cos^4 \varphi+\sin2\varphi,\\ 
		b&=\frac{N}{4}(g_{12}-2g)\cos^2 \varphi , \\ 
		c&=\frac{N}{4}[g+g_{12}-(3g+2g_{12})\cos^2 \varphi], \\ 
		d&=\frac{N}{8}[g+g_{12}+(g-g_{12})\cos^2 \varphi], \\ 
		e&=-\frac{gN}{4}\cos^2 \varphi , \\ 
		f&=\frac{N}{8}[g+g_{12}-(3g+g_{12})\cos^2 \varphi].
	\end{split}
\end{align}
Note that its characteristic equation is an eighth-degree polynomial equation 
in even powers of $\omega$ and we thus obtain its eigenvalues $\pm\omega_{\textrm{in},i}$ with $i=1,\dots,4$, 
\begin{align}
	\label{BdGOPomegam}
	\omega_{\textrm{in},1}&=\sqrt{r_1-2\sqrt{s_1}},\\
	\omega_{\textrm{in},2}&=\sqrt{r_1+2\sqrt{s_1}}, \\ 
	\omega_{\textrm{in},3}&=\sqrt{r_2-2\sqrt{s_2}}, \\ 
	\omega_{\textrm{in},4}&=\sqrt{r_2+2\sqrt{s_2}},
\end{align}
with
\begin{align}
	\label{BdGOPomegamrs}
	r_1=&1+(a-b)^2+(b+c)^2-(d-e)^2-(e+f)^2,\\
	r_2=&1+(a+b)^2+(b-c)^2-(d+e)^2-(e-f)^2,\\ 
	s_1=&b^4-2b^3(a-c)-4ab^2c+a^2[1+(b+c)^2]-2c(a-b)(d-e)(e+f)+[(d-e)^2-1](e+f)^2 \\ \nonumber
	&+b^2[1+c^2+2(d-e)(e+f)]-2ab[1+c^2+(d-e)(e+f)],\\ 
	s_2=&b^4+2b^3(a-c)-4ab^2c+a^2[1+(b-c)^2]+2c(a+b)(d+e)(e-f)+[(d+e)^2-1](e-f)^2 \\ \nonumber
	&+b^2[1+c^2-2(d+e)(e-f)]+2ab[1+c^2-(d+e)(e-f)].
\end{align}
As shown in Supplementary Fig.~\ref{Fig_omega_m}b, the analytical result of $\text{Im}(\omega_{\textrm{in}})$ 
agrees with that of the full BdG matrix.

\section{Supplementary Note 6. The continuous model}

\subsection{A. The Bloch states through gauge transformation}
The continuous model $H^{\mathrm{lin}}_{\mathrm{c}}(x,\theta)$ is invariant under translation by 
$2a$, i.e., $H^{\mathrm{lin}}_{\mathrm{c}}(x+2a,\theta)=H^{\mathrm{lin}}_{\mathrm{c}}(x,\theta)$.
We find that its two lowest Bloch bands intersect (see the solid and dashed red lines 
in Supplementary Fig.~\ref{Fig_TB_con} in the interval $[0,\pi/a]$), which contrasts 
with the commonly studied Thouless pump where either a single band or multiply isolated 
bands are occupied. To address this issue, in this subsection, we will demonstrate that the original Hamiltonian
can be transformed into one that is invariant under translation by $a$, 
thanks to a symmetry. 
In the subsequent subsection, we will prove that 
the linear Thouless pump for the continuous model exhibits a displacement of $C a$ 
per particle during a pump cycle, where $C$ is the corresponding Chern number.

Consider a system with the length $L=M a$. Let $\phi(x,\theta)$ be an 
instantaneous eigenstate of $H_{c}^{\textrm{lin}}(x,\theta)$ at a fixed $\theta$, 
i.e., 
\begin{equation} \label{Eq:H_c}
	H_{\mathrm{c}}^{\textrm{lin}}(x,\theta) \phi(x,\theta) = E(\theta) \phi(x,\theta).
\end{equation}
As the Wannier functions of $H_{0}(x)$ constitute a basis for the Hilbert space, we can write 
$\phi(x,\thr)$ as a linear combination of the Wannier functions at each $\theta$, 
\begin{equation} \label{Eq:phi}
	\phi(x,\thr)=\left(\begin{array}{c}
		\phi_{1}(x,\thr)\\
		\phi_{2}(x,\thr)
	\end{array}\right)=\sum_{nj}W_{nj}(x)\left(\begin{array}{c}
		(-1)^{j}c_{1nj}(\theta)\\
		c_{2nj}(\theta)
	\end{array}\right),
\end{equation}
where 
$W_{nj}(x)$ is the $j$th Wannier function of the $n$th Bloch band of $H_0(x)$, and
given a Wannier function $W_n (x)$ for $j=0$ from the $n$th Bloch band, 
$W_{nj}(x)$ is obtained as $W_n (x-ja)$~\cite{Wannier}. 
Here, the Wannier functions are not required to be maximally localized, 
and $c_{\sigma nj}(\theta)$ ($\sigma=1,2$)
are weights taking complex values. The phase $(-1)^j$ is applied, as explained below.
For notation simplicity, we drop the parameter $\theta$ in the following.
Substituting Eq.~(\ref{Eq:phi}) into Eq.~(\ref{Eq:H_c}) gives rise to the following equation
for the weights of $c_{\sigma nj}$,
\begin{equation}
	\sum_{\sigma^{\prime}n^{\prime}j^{\prime}}H_{\sigma nj,\sigma^{\prime}n^{\prime}j^{\prime}}c_{\sigma^{\prime}n^{\prime}j^{\prime}}=Ec_{\sigma nj},
\end{equation}
where 
\begin{equation}
	H_{\sigma nj,\sigma^{\prime}n^{\prime}j^{\prime}}= 
	(-1)^{j \sigma+j^\prime \sigma^\prime }
	\int {\rm d}xW_{nj}(x)[H_{\mathrm{c}}^{\textrm{lin}}(x)]_{\sigma\sigma^{\prime}}W_{n^{\prime}j^{\prime}}(x).
\end{equation}
The Hamiltonian $H$ is thus the representation of $H_{\mathrm{c}}^{\textrm{lin}}(x)$ relative to a new basis 
aided by the gauge transformation.
Since the Hamiltonian $H_{\mathrm{c}}^{\textrm{lin}}(x)$ preserves the $T_a \sigma_z$ symmetry   
where $T_a$ is the translation operator by $a$, i.e., 
$T_a \sigma_z H_{\mathrm{c}}^{\textrm{lin}}(x) (T_a \sigma_z)^{-1}=H_{\mathrm{c}}^{\textrm{lin}}(x)$, 
we can prove that $H$ is translation invariant under one lattice site through the following steps, 
\begin{eqnarray}
	H_{\sigma nj+1,\sigma^{\prime}n^{\prime}j^{\prime}+1} 
	& = &\int {\rm d}xW_{nj+1}(x)(-1)^{(j+1)\sigma}[H_{\mathrm{c}}^{\textrm{lin}}(x)]_{\sigma\sigma^{\prime}}W_{n^{\prime}j^{\prime}+1}(x)(-1)^{(j^{\prime}+1)\sigma^{\prime}}\\
	& = &\int {\rm d}xW_{nj}(x)(-1)^{(j+1)\sigma}[H_{\mathrm{c}}^{\textrm{lin}}(x+a)]_{\sigma\sigma^{\prime}}W_{n^{\prime}j^{\prime}}(x)(-1)^{(j^{\prime}+1)\sigma^{\prime}}\\
	& = &H_{\sigma nj,\sigma^{\prime}n^{\prime}j^{\prime}}.
\end{eqnarray}
In the proof, we have used the fact that 
$(-1)^{\sigma+\sigma^\prime}[H_{\mathrm{c}}^{\textrm{lin}}(x+a)]_{\sigma \sigma^\prime}=[H_{\mathrm{c}}^{\textrm{lin}}(x)]_{\sigma \sigma^\prime}$
enforced by the $T_a \sigma_z$ symmetry.
The consequence is that the eigenstates
of $H$ are Bloch states, i.e., $Hc^{(lk)}=\epsilon_{lk}c^{(lk)}$
where the entries of $c^{(lk)}$ are given by $c_{\sigma nj}^{(lk)}=\frac{1}{\sqrt{M}}e^{ikja}\tilde{c}_{\sigma n}^{(lk)}$
with $k\in[0,2\pi/a]$ for the $l$th band of $H$ and $\sum_{\sigma n}|\tilde{c}_{\sigma n}^{(lk)}|^{2}=1$. 
The corresponding eigenstate $\phi(x)$ is
\begin{equation}
	\phi_{lk}(x)=\frac{1}{\sqrt{M}}\sum_{nj}W_{nj}(x)e^{ikja}\left(\begin{array}{c}
		(-1)^{j}\tilde{c}_{1n}^{(l\kr)}\\
		\tilde{c}_{2n}^{(l\kr)}
	\end{array}\right).
\end{equation}

\begin{figure}
	\includegraphics[width = 0.3\linewidth]{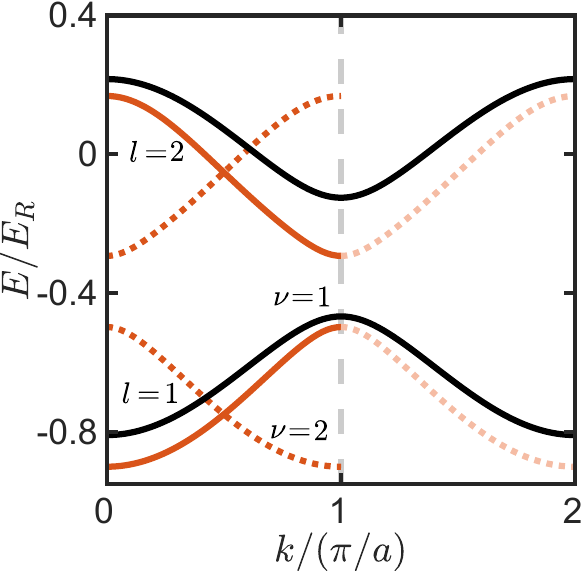}
	\caption{
		The eigenenergies (red lines) of the lowest two Bloch bands of 
		the Hamiltonian $H^{\mathrm{lin}}_{\mathrm{c}}(x,\theta)$
		and those (black lines) of the tight-binding Hamiltonian 
		$\tilde{H}(k,\theta)$ in Eq.~(\ref{Hamk}) at $\theta=0$ with respect to momentum $k$.
		Since $H^{\mathrm{lin}}_{\mathrm{c}}(x,\theta)$ is periodic with a period of $2a$,
		its eigenenergies are indexed by the band index $l$, the subband index $\nu=1,2$
		and momentum $k\in[0,\pi/a]$. Due to a symmetry, the Bloch state can also be identified 
		with momentum within $[0,2\pi/a]$. For example, the subband (dotted red line) with $\nu=2$ 
		can be shifted into
		the region $[\pi/a,2\pi/a]$ (dotted light red line), forming a single band together with the 
		subband with $\nu=1$. 
	}
	\label{Fig_TB_con}
\end{figure}

We see that although the Hamiltonian $H_{c}^{\textrm{lin}}$ is not
invariant under a translation of $a$, the Bloch state can also be
characterized by the momentum $k\in[0,2\pi/a]$. 
Given that $W_{nj}(x)=\frac{1}{\sqrt{M}}\sum_{k}e^{-ik ja}\psi_{nk}^{(0)}(x)$
where $\psi_{nk}^{(0)}(x)$ is a Bloch state of $H_{0}$ in the $n$th
band at momentum $k$,
we obtain 
\begin{equation}
	\phi_{lk}(x)=\sum_{n}\left(\begin{array}{c}
		\psi_{n,k+\pi/a}^{(0)}(x)\tilde{c}_{1n}^{(l\kr)}\\
		\psi_{nk}^{(0)}(x)\tilde{c}_{2n}^{(l\kr)}
	\end{array}\right)=\sum_{n}\left(\begin{array}{c}
		e^{i(k+\pi/a)x}u_{n,k+\pi/a}^{(0)}(x)\tilde{c}_{1n}^{(l\kr)}\\
		e^{ikx}u_{nk}^{(0)}(x)\tilde{c}_{2n}^{(l\kr)}
	\end{array}\right)=e^{ikx}f_{lk}(x)
\end{equation}
where $f_{lk}(x)=\sum_{n}\left(\begin{array}{c}
	e^{i(\pi/a)x}u_{nk+\pi/a}^{(0)}(x)\tilde{c}_{1n}^{(l\kr)}\\
	u_{nk}^{(0)}(x)\tilde{c}_{2n}^{(l\kr)}
\end{array}\right)$.
In addition, since
$H_{\mathrm{c}}^{\textrm{lin}}(x)\phi_{lk}(x)=\epsilon_{lk}\phi_{lk}(x)$, we
have $H_{\mathrm{c}}^{\textrm{lin}}(k)f_{lk}(x)=\epsilon_{lk}f_{lk}(x)$, where
\begin{equation}
	H_{\mathrm{c}}^{\textrm{lin}}(k)=\frac{(p_{x}+\hbar k)^{2}}{2m}-V_{x}\cos^{2}(k_{R}x)+h_z(\theta)\sigma_{z}+V_{\textrm{so} }(x,\theta)\sigma_{x}.
\end{equation}

We have seen that the Bloch state $\phi_{lk}(x)$ of $H_{\mathrm{c}}^{\textrm{lin}}(x)$ can be expressed in terms of 
the eigenvectors $c^{(lk)}$ of $H$ characterized by the momentum $k$ within the interval $[0,2\pi/a]$. 
Given that $H_{\mathrm{c}}^{\textrm{lin}}(x+2a)=H_{\mathrm{c}}^{\textrm{lin}}(x)$,  
the Bloch states can alternatively be described by the momentum $k^\prime$ within the interval $[0,\pi/a]$,
as illustrated by the red lines in Supplementary Fig.~\ref{Fig_TB_con}. For clarity, the states are denoted as  
$\chi_{l,\nu, k^\prime}(x)=e^{ik^\prime x} u_{l,\nu, k^\prime }(x)$, where $\nu=1,2$ 
correspond to the two subbands (see Supplementary Fig.~\ref{Fig_TB_con}). Since both $\phi_{lk}(x)$ and $\chi_{l,\nu, k^\prime}(x)$
are Bloch states of $H_{\mathrm{c}}^{\textrm{lin}}(x)$, they are equivalent. 
Therefore, let $\phi_{lk}(x)=\chi_{l,1, k}(x)$ when $k\in[0,\pi/a]$.
For $k\in(\pi/a,2\pi/a]$, we have $\phi_{lk}(x)=e^{ikx}f_{lk}(x)=e^{i(k-\pi /a)x}f_{lk}(x)e^{i\pi x/a}$,
which equals $\chi_{l,2,k-\pi/a}=e^{i (k-\pi /a)x}u_{l,2,k-\pi/a}(x)$, resulting in
$u_{l,2,k-\pi/a}(x)=f_{lk}(x)e^{i\pi x/a}$.

\subsection{B. The linear Thouless pumping}
We are now in a position to derive the displacement of atoms for the continuous
model as we tune a system parameter $\theta$ from $\theta(t=0)=0$ to $\theta(t)$.
Consider a cloud of fermionic atoms occupying the $l$th band of the model.  At each momentum, a
state evolves as $|\psi_{lk}(t)\rangle$ governed by the time-dependent
Schr\"{o}dinger equation:
\begin{equation}
	i\hbar\partial_{t}|\psi_{lk}(t)\rangle=H_{\mathrm{c}}^{\textrm{lin}}|\psi_{lk}(t)\rangle.
\end{equation}
Since the velocity operator is 
\begin{equation}
	v=-\frac{i}{\hbar}[x,H_{\mathrm{c}}^{\textrm{lin}}(x)]=p_{x}/m,
\end{equation}
the average displacement of the cloud per an atom at time $t$ is 
\begin{eqnarray} \label{Avg-Dis-Eq}
	x_{c}(t) & =&\frac{1}{L/a}\frac{L}{2\pi}\int_{0}^{2\pi/a}{\rm d}k\int_{0}^{t}\langle\psi_{lk}(\tau)|v|\psi_{lk}(\tau)\rangle {\rm d}\tau\\
	& =&\frac{a}{2\pi}\frac{1}{m}\int_{0}^{2\pi/a}{\rm d}k\int_{0}^{t}\langle\psi_{lk}(\tau)|p_{x}|\psi_{lk}(\tau)\rangle {\rm d}\tau.
\end{eqnarray}
If $\theta$ is varied very slowly, then we have~\cite{xiao2010berry}
\begin{equation}
	|\psi_{lk}(t)\rangle=e^{-i\frac{1}{\hbar}\int_{0}^{t}\epsilon_{lk}{\rm d}\tau}\left[|\phi_{lk}(t)\rangle+\sum_{l^{\prime}\neq l}\frac{i\hbar}{\epsilon_{l^{\prime}k(t)}-\epsilon_{lk(t)}}|\phi_{l^{\prime}k}(t)\rangle\langle\phi_{l^{\prime}k}(t)|\partial_{t}\phi_{lk}(t)\rangle \right].
\end{equation}
Based on this equation, the expectation value of the momentum operator $p_x$ at time $t$ is
\begin{eqnarray} \label{Mom-Eq}
	&&\langle\psi_{lk}(t)|p_{x}|\psi_{lk}(t)\rangle  \nonumber \\
	& = &\langle\phi_{lk}(t)|p_{x}|\phi_{lk}(t)\rangle+\sum_{l^{\prime}\neq l}\frac{i\hbar}{\epsilon_{l^{\prime}k(t)}-\epsilon_{lk(t)}}\left[\langle\phi_{lk}(t)|p_{x}|\phi_{l^{\prime}k}(t)\rangle\langle\phi_{l^{\prime}k}(t)|\partial_{t}\phi_{lk}(t)\rangle \right. \nonumber \\
	&&\left.-\langle\partial_{t}\phi_{lk}(t)|\phi_{l^{\prime}k}(t)\rangle\langle\phi_{l^{\prime}k}(t)|p_{x}|\phi_{lk}(t)\rangle \right] \nonumber \\
	&=& \frac{m}{\hbar}\frac{\partial\epsilon_{lk}}{\partial k} +m\Omega_{l}(k,t),
\end{eqnarray}
where $\Omega_{l}(k,t)= i[\langle \partial_{t}f_{lk}(t) |\partial_{k}f_{lk}(t)\rangle-\text{c.c.}]=\partial_{t}A_{l,k}-\partial_{k}A_{l,t}$
is the Berry curvature with $A_{l,t}=i\langle f_{lk}(t)|\partial_{t}f_{lk}(t)\rangle$
being the Berry connection. In the derivation, we have used the fact that
\begin{equation}
	\langle f_{l^{\prime}k}(t)|\frac{\partial H_{\mathrm{c}}^{\textrm{lin}}(k)}{\partial k}|f_{lk}(t)\rangle=\langle f_{l^{\prime}k}(t)|\partial_{k}f_{lk}(t)\rangle(\epsilon_{l^{\prime}k(t)}-\epsilon_{lk(t)}).
\end{equation}
Substituting Eq.~(\ref{Mom-Eq}) into Eq.~(\ref{Avg-Dis-Eq}) leads to
\begin{equation}
	x_{c}(t)=\frac{a}{2\pi}\int_{0}^{t} {\rm d}\tau \int_{0}^{2\pi/a}{\rm d}k\Omega_{l}(k,\tau)=\frac{a}{2\pi}\int_{0}^{t}\frac{\partial\gamma_{l}(\tau)}{\partial\tau}{\rm d}\tau,
\end{equation}
where $\gamma_{l}(\tau)=\int_{0}^{2\pi/a}{\rm d}kA_{l,k}(\tau)$ is the
Zak phase along the path of $k$ running from $0$ to $2\pi/a$ at a fixed $\tau$.
Consequently, the average displacement over a pump cycle $T$ is $x_{c}=a C_l$, where 
$C_l=[\gamma_l(T)-\gamma_l(0)]/(2\pi)$ is the Chern number for the $l$th band.
Here we require that as $\tau$ varies, $\gamma_l(\tau)$ changes continuously. 
To numerically evaluate $\gamma_{l}(\tau)$, one can divide the Brillouin zone into $N_0$ parts so that 
the Zak phase can calculated via the following formula~\cite{resta2000manifestations}
\begin{equation}
	\gamma(\tau)=i\ln \left[\langle f_{l,0}(\tau)|f_{l,\Delta k}(\tau)\rangle\langle f_{l,\Delta k}(\tau)|f_{l,2\Delta k}(\tau)\rangle\langle f_{l,2\Delta k}(\tau)|\dots\langle 
	f_{l,(2\pi/a)-\Delta k}(\tau)|f_{l,2\pi/a}(\tau)\rangle \right],
\end{equation}
where $|f_{l,2\pi}(\tau)\rangle=|f_{l,0}(\tau)\rangle$ and $\Delta k=2\pi/(a N_0)$. 

\subsection{C. Tight-binding model on the lowest Bloch band}
In this subsection, we will follow Refs.~\cite{liu2014realization,yangPRB2018, 
	wuSci2016,xuPRA2016_Dirac,xuPRA2016_typeII} to derive the tight-binding model by projecting the continuous Hamiltonian 
in Eq. (4) in the main text to the lowest band of $H_0(x)$. Let us first write 
$\psi_\sigma(x,t)=e^{-i \mu(\theta) t} \chi_\sigma(x,\theta)$ where $\mu(\theta)$ is the chemical potential and $\chi_\sigma(x,\theta)$
satisfies the following time-independent GP equation:
\begin{equation}\label{Eq:GP}
	\mu \chi_\sigma (x,\theta)= 
	[\tilde{H}_{\mathrm{c}}^{\textrm{lin} }(x,\theta)\chi (x,\theta)]_\sigma 
	+g|\chi_{\sigma}(x,\theta)|^2\chi_{\sigma}(x,\theta).
\end{equation}
Here, $\chi(x,\theta)=(\chi_1(x,\theta),\chi_2(x,\theta))^T$.
To derive the tight-binding model, 
we approximate $\chi(x,\theta)$ by considering only contributions from 
the Wannier functions $W_{nj}(x)$ of the lowest band of $H_{0}(x)$ with $n=1$. Specifically, 
\begin{equation} \label{Eq:Wann}
	\chi(x,\theta) \approx \sum_{j}W_{1j}(x)\left(\begin{array}{c}
		(-1)^{j}\chi_{1j}(\theta)\\
		\chi_{2j}(\theta)
	\end{array}\right),
\end{equation}
where $\chi_{1j}(\theta)$ and $\chi_{2j}(\theta)$ are coefficients, and
the phase $(-1)^j$ is applied similar to 
the gauge transformation employed in Subsection A.
For notation simplicity, we drop the parameter $\theta$ in the following.
Substituting Eq.~(\ref{Eq:Wann}) into Eq.~(\ref{Eq:GP}) and keeping only the onsite and nearest-neighbor hopping,
we obtain
\begin{equation}
	\mu \chi_{\sigma j}= 
	[\tilde{H}_{\textrm{TB} }\chi_j]_\sigma +
	g\sum_{j_1,j_2,j_3}W_{jj_1 j_2 j_3} \chi_{\sigma j_1}^* \chi_{\sigma j_2} \chi_{\sigma j_3}.
\end{equation}
where $\chi_j=(\chi_{1j},\chi_{2j})^T$ and $W_{jj_1 j_2 j_3}=\int {\rm d}x W_{1j}^*(x) W_{1 j_1}^*(x) W_{1 j_2}(x) W_{1 j_3}(x)$.
The approximate tight-binding Hamiltonian is
\begin{equation}\label{TB2}
	{\tilde{H}}_{\textrm{TB} }=\sum_j \left[t_0{a}^\dagger_{j}{a}_{j} 
	+(t_1{a}^\dagger_{j}\sigma_z{a}_{j+1}+{\rm H.c.} )
	+h_z(\theta){a}^\dagger_{j}\sigma_z{a}_{j} 
	+(i {t}_s {a}^\dagger_{j}\sigma_y {a}_{j+1}+{\rm H.c.})
	+ {t}_c\sin(\theta) {a}^\dagger_{j}\sigma_x {a}_{j}
	\right],
\end{equation}  
where ${a}^\dagger_{j}=(|1 j\rangle, |2 j\rangle)$ and
${a}_{j}=(\langle 1j |, \langle 2j |)^T$ with $|\sigma j\rangle$ describing 
the $\sigma$th degree of freedom in the $j$th unit cell.
The hopping parameters are given by
$t_0=\int {\rm d} xW_{1j}(x) H_0(x)W_{1j}(x)$,
$t_1=-\int {\rm d} xW_{1j}(x) H_0(x)W_{1,j+1}(x) $,
$t_s=(-1)^j t_s^{j,j+1}$ with
$t_s^{j,j+1}=V_s \int {\rm d} x W_{1j}(x) \sin(k_Rx) W_{1,j+1}(x)$,
and $t_c=(-1)^j t_c^{j,j}$ with $t_c^{j,j}=V_c \int {\rm d} xW_{1j}(x) \cos(k_Rx)W_{1j}(x)$.
Note that we do not have terms with $t_s^{j,j}$ and $t_c^{j,j+1}$, which are forced to be zero due
to the fact that $W_{10}(x)$ is an even function.
By Fourier transformation, we transform the Hamiltonian in Eq.~(\ref{TB2}) into the 
momentum-space form:
\begin{equation}\label{Hamk}
	\tilde{H}(k,\theta)=t_0\sigma_0+[h_z(\theta)+2t_1\cos(k)]\sigma_z-2 {t}_s\sin(k)\sigma_y
	+ {t}_c\sin(\theta)\sigma_x,
\end{equation}  
which is the same as the Hamiltonian $H^{\mathrm{lin}}$ in Eq. (2) in the main text except a 
constant term $t_0\sigma_0$ by setting 
$2t_1=J_1$, $2 t_s=-J_1^\prime$, $t_c \sin \theta=J_2$, and $h_z=m_z$.

In Supplementary Fig.~\ref{Fig_TB_con}, we compare the energy spectrum (black lines) 
of $\tilde{H}(k,\theta)$ at $\theta=0$ with that (red lines) of the continuous Hamiltonian $H_{\mathrm{c}}^{\mathrm{lin}}$, 
demonstrating their similarity. In addition, 
the Chern number of the lowest band of $\tilde{H}(k,\theta)$ and the continuous model is identical. 

For the nonlinear part, we approximate $W_{jj_1 j_2 j_3}$ as 
$W_{jjjj} \delta_{j j_1} \delta_{j j_2} \delta_{j j_3}$ by considering the 
largest contribution and neglecting all other terms. We find that the soliton coming 
from the lowest band with repulsive interaction ($g>0$) in the tight-binding model 
exhibits pumping with twice the linear Chern number. In contrast, the pumping in the continuous 
model equals the linear Chern number, suggesting that we may oversimplify the nonlinear effects.

\end{widetext}

\end{document}